
\magnification=1200
\hoffset=1truecm
%
%
\def\uplrarrow#1{\raise1.5ex\hbox{$\leftrightarrow$}\mkern-16.5mu #1}
\def\bx#1#2{\vcenter{\hrule \hbox{\vrule height #2in \kern #1\vrule}\hrule}}

\def\tr{\,{\hbox{tr}}\,}

\def\det{\,{\hbox{det}}\,}

\def\mod{\,{\hbox{mod}}\,}

\def\squiggle#1{\lower1.5ex\hbox{$\sim$}\mkern-14mu #1}

\def\thru#1{\mathrel{\mathop{#1\!\!\!\!/}}}

\def\narrower{\advance\leftskip by\parindent \advance\rightskip by\parindent}

\def\figcaptr#1#2{\narrower\smallskip\noindent Fig. #1
                \quad #2\smallskip}

\def\mbox#1#2{\vcenter{\hrule width#1in\hbox{\vrule height#2in
   \hskip#1in\vrule height#2in}\hrule width#1in}}
\def\eqsquare #1:#2:{\vcenter{\hrule width#1\hbox{\vrule height#2
   \hskip#1\vrule height#2}\hrule width#1}}
\def\inbox#1#2#3{\vcenter to #2in{\vfil\hbox to #1in{$$\hfil#3\hfil$$}\vfil}}
\def\strutdepth{\dp\strutbox}
\def\marbul{\strut\vadjust{\kern-\strutdepth\specialbul}}
\def\specialbul{\vtop to \strutdepth{
    \baselineskip\strutdepth\vss\llap{$\bullet$\qquad}\null}}
\def\Bcomma{\lower6pt\hbox{$,$}}    
\def\bcomma{\lower3pt\hbox{$,$}}    

\def\updots{\mathinner{\mskip 1mu\raise 1pt\hbox{.}
    \mskip 2mu\raise 4pt\hbox{.}\mskip 2mu
    \raise 7pt\vbox{\kern 7pt\hbox{.}}\mskip 1mu}}

\def\pmb#1{\setbox0=\hbox{#1}%
     \kern-.025em\copy0\kern-\wd0
     \kern.05em\copy0\kern-\wd0
     \kern-.025em\raise.0433em\box0}

\def\R{\;\,R\!\!\!\!\!\!\!\> R\,\,}
\def\Z{\;\>Z\!\!\!\!\!\! Z\;\;}

\def\1{\;1\!\!\!\! 1\;}
\def\eg{{\it e.g.}}
\def\ie{{\it i.e.}}

\def\m@th{\mathsurround=0pt}
\def\upsquarefill{$\m@th\bracelu\leaders\vrule\hfill\braceru$}
\def\ope#1{\mathop{\vtop{\ialign{##\crcr
     $\hfil\displaystyle{#1}\hfil$\crcr\noalign{\kern3pt\nointerlineskip}
     \kern4pt\upsquarefill\kern4pt\crcr\noalign{\kern3pt}}}}\limits}
\def\Int#1#2{\int\!d^{#1}{#2}\,}
\def\lsim{\mathrel{\rlap{\lower4pt\hbox{\hskip1pt$\sim$}}
    \raise1pt\hbox{$<$}}}         
\def\gsim{\mathrel{\rlap{\lower4pt\hbox{\hskip1pt$\sim$}}
    \raise1pt\hbox{$>$}}}         

\magnification=1200
\vsize=7.5in
\hsize=5in
\pageno=0
\tolerance=10000

\def\dotx{{\dot{\vec x}}}

\hfuzz=5pt
\baselineskip 12pt plus 2pt minus 2pt
\hfill DFTT 93/33

\hfill hep-th/9307047

\hfill July 1993
\bigskip
\centerline{\bf RELATIVISTIC QUANTUM THEORY}
\smallskip
\centerline{\bf WITH FRACTIONAL SPIN AND STATISTICS}
\vskip 36pt\centerline{Stefano Forte }
\vskip 12pt
\centerline{\it I.N.F.N., Sezione di Torino }

\centerline{\it via P.~Giuria 1,
I-10125, Torino, Italy }
\vskip 1.5in
\centerline{Lectures presented at}
\medskip
\centerline{VIII J.~A.~Swieca summer
school ``Particles and Fields''}
\centerline{Campos do Jord\~ao, Brazil, January 1993}
\smallskip
\centerline{\it to be published in the proceedings}
\centerline{\it O.~Eboli and V.~Rivelles, Ed.s (World Scientific, 1993)}
\bigskip
\centerline{\it and}
\bigskip
\centerline{Workshop on ``Algebraic Topology and Anyon
Superconductivity''}
\centerline{Pozna\'n, Poland, May 1993}
\vfill
\eject
\eject
\baselineskip 16pt plus 3pt minus 3pt
\centerline{\bf CONTENTS}
\bigskip
\item{I.}\quad Why relativistic anyons? \dotfill 2
\bigskip
\item{II.}\quad From the nonrelativistic to the
relativistic theory  \dotfill 6
\medskip
\itemitem{II.1}\quad The path integral approach \dotfill 7
\itemitem{II.2}\quad Covariant formulation and Chern-Simons theory\dotfill 11
\item{}\quad {\it Mathematical digression}: The Lorentz
group in 2+1 dimensions
\dotfill 15
\bigskip
\item{III.}\quad Spin without spinors\dotfill 19
\medskip
\itemitem{III.1}\quad Path integrals for spin \dotfill 19
\item{}\quad {\it Mathematical digression}: The Poincar\'e
group and point particles
\item{}\qquad in 2+1 dimensions\dotfill 27
\itemitem{III.2}\quad The relativistic spinning particle \dotfill 29
\bigskip
\item{IV.}\quad Point particles with generic spin \dotfill 36
\medskip
\itemitem{IV.1}\quad The Hopf action and the spinning particle\dotfill 36
\item{}\quad {\it Mathematical digression}: The
Hopf fibration and the Dirac monopole \dotfill 41
\itemitem{\quad IV.2}\quad Path integral and multivalued
relativistic wave function \dotfill 44
\bigskip
\item{V.}\quad Relativistic field theory\dotfill 58
\medskip
\itemitem{V.1}\quad The Klein-Gordon-Chern-Simons theory
\dotfill 58
\itemitem{V.2}\quad The operator cocycle approach\dotfill 63
\bigskip
\item{VI.}\quad Open Problems\dotfill 74
\bigskip
\item{}\quad References\dotfill 75
\vfill
\eject
\input harvmac
\baselineskip 16pt plus 2pt minus 2pt
\newsec{\bf WHY RELATIVISTIC ANYONS?}

Quantum mechanical particles can have only integer or half-integer spin, and
their statistics must be accordingly Bose or Fermi --- this was the standard
wisdom until a few years ago. Now
we know
that the truth of this statement depends on the number of space-time
dimensions:
it fails if the space dimension is less than three.
Both spin and statistics are physically not very interesting in one dimension.
On the other hand, many realistic systems have effectively only two
dimensions, and the possibility of generalizing spin and statistics in the
planar case seems intriguing and attractive.
 However, despite the excitement it has generated,
no really solid application
to realistic physical systems of this discovery has been
found yet. Applications to the formulation of a theory of high-$T_c$
superconductivity are still at best tentative. Even in the case of the
fractional Quantum Hall effect, where the best available theory seems to
display this phenomenon, fractional spin\foot{ It has become customary
to call generalized spin and statistics
{\it fractional statistics}. This is somewhat
of a misnomer, since in fact {\it all} real values of the spin and statistics
are possible in two spatial dimensions, and not only rational ones as
``fractional'' would seem to suggest. We shall however stick to this now
conventional nomenclature. Excitations with generic spin are often also called
{\it anyons} (as opposed to bosons and fermions).
Fractional statistics should not be confused with {\it parastatistics}.
Particles with parastatistics have ordinary (\ie, integer or half-integer)
spin, but satisfy a modified exclusion principle. Particles with fractional
statistics satisfy the  ordinary exclusion principle (\ie, no two particles can
occupy the same quantum state), but have generic spin.}
is hardly more than an {\it a
posteriori} feature of the theory, rather than providing a physical explanation
of the observed effect.

However, quantum mechanics with generic spin and statistics offers the
possibility of rediscovering some of the basic ideas of quantum mechanics in a
new light, whatever its phenomenological success in condensed
matter physics. Perhaps, new, purely
quantum mechanical effects may be found. Certainly,
a more general framework to
understand the quantum mechanics of spin appears.

But then, the dynamics of spin is nontrivial only in the relativistic case,
as
we can see by looking at the simplest physical example, that of a free
spinning particle. Whereas in non-relativistic quantum mechanics the wave
function is just the tensor product of a spatial wave function $\psi(\vec x)$
times
a spin wave function $u$,
in the relativistic case the spin and translational
degrees of freedom are coupled dynamically by the Dirac equation
$(\thru \partial + m)\psi(x)=0$ which must
be satisfied by the spinor wave function $\psi(x)$.

\goodbreak
\bigskip{\narrower
\noindent{\it
Example: the Hausdorff dimension of paths for a free particle}
\smallskip
\nobreak
The dynamical coupling of spin and translational degrees of freedom
has observable, physical effects even for a single free
particle.
This can be seen by computing the Hausdorff dimension of the  paths
traversed by a quantum mechanical free particle, \ie,
the scaling law which gives the typical length $L$
of a path traversed by a quantum particle
in terms of the distance $R$ between the initial and final points of the path.
Indeed, it can be shown that the paths that contribute to the
quantum propagation amplitude, \ie\ to the path integral for the particle
propagation satisfy the scaling law
\eqn\scal
{R=L^{1/d_H},}
where the Hausdorff dimension $d_H$ takes a well-defined value.
This value turns out to be different for spinning and spinless particles.

We can understand that
via an argument due to Polyakov,\ref\polya{
A.~M.~Polyakov, in ``Fields, Strings and Critical Phenomena'',
Les Houches summer school proceedings, session XLIX, edited by
E.~Br\'ezin and J.~Zinn-Justin (North Holland, Amsterdam, 1989);
\semi  ``Gauge Fields and Strings'' (Harwood, Chur, 1987).}
by considering the quantum propagator $
K(x^\prime,x)=\langle
x^\prime|x\rangle$ that connects two points $x$, $x^\prime$ such that
$|x^\prime-x|=R$. This is
just the Fourier transform of the usual momentum-space
propagator $K(p)$:
\eqn\prop
{\langle
x^\prime|x\rangle=\int\!d\vec p \ e^{i p(x^\prime-x)}\,K(p)}
where, for free bosons and fermions, respectively
\eqn\pbf
{\hbox{\it Bosons:}\>\> K(p)=K^B(p)={1\over p^2+m^2}\>\>\qquad
\hbox{\it Fermions:}\>\> K(p)=K^F(p)={1\over \thru p+m},}
\ie, the propagator is respectively equal to
the Klein-Gordon and the Dirac one.
The propagators \prop\ may be rewritten as
\eqn\pint
{\eqalign{K^B(x^\prime,x)&=\int_0^\infty\!dL\,e^{-m^2L}\int\!d\vec p\,
e^{ip(x^\prime-x)-Lp^2}\cr
K^F(x^\prime,x)&=\int_0^\infty\!dL\,e^{-mL}\int\!d\vec p\,
e^{ip(x^\prime-x)-L\thru p}.\cr}}

The integration over $L$ in Eq,\pint\ can be actually viewed
as the integration over the lengths of the paths
which contribute to the propagator.
That is, if we express $K(x^\prime,x)$ as a Feynman path integral,
and we define
$K(L;x^\prime,x)$ by the relation
\eqn\pathl
{K(x^\prime,x)=\int\!dL\, e^{-Lm^{d_H}} K(L;x^\prime,x),}
where $d_H=2$ for bosons and $d_H=1$ for fermions,
then $K(L;x^\prime,x)$ is the contribution to the propagator
$K(x^\prime,x)$ which is obtained restricting the Feynman
sum over paths to paths of
length $L$. A rigorous proof of this goes beyond our point; the simplest way of
seeing that this is the case
is to notice that the propagator $K(x^\prime,x)$ provides a
solution to the Klein-Gordon and Dirac equations in the cases of bosons and
fermions respectively. All the solution of these equations
can be obtained from the solutions to the equation
\eqn\diff
{{\partial\over
\partial s}\phi=\hat O
\phi}
where $\hat O=\partial_\mu\partial^\mu$ and $\hat O=\thru\partial$ in the
respective cases, by projecting out the solutions $\phi$ that satisfy
${\partial\over
\partial s}\phi=m^{d_H}\phi$. But if we identify $s$ with proper time, then
the solutions to Eq.\diff\ are given by the Feynman sum over paths with fixed
length $L$, while $K(x^\prime, x)$
is given by Eq.\pathl\ with $L$ identified   with the path length.

Inspection of Eq.\pint\ then reveals immediately that the bulk of the
contribution to the propagator $K(x^\prime,x)$ comes from paths which satisfy
\eqn\scala{
L\sim{1\over p^{d_H}},}
which gives immediately Eq.\scal\ because $L\sim{1\over R}$.
We see thus  that indeed the
scaling (Hausdorff) dimension $d_H$ equals two for bosonic paths, and one
for fermionic ones. In other words the value of spin modifies the (fractal)
properties of the paths traversed by a free quantum particle: spin makes its
presence felt even in the absence of any explicit spin dependent potential
or coupling to other particles.\bigskip}

\noindent
But if it is only in a relativistic theory that
the spin dynamics is nontrivial,
we are naturally lead to ask,
does the dynamics of particles and fields with fractional
spin and statistics admit a
relativistic formulation? And if yes, how will it look like?
It is the purpose of the present lectures to answer these questions.

In order to make
our treatment as self-contained as possible, we shall start by reviewing
in Sect.II the
nonrelativistic theory, concentrating on the path-integral approach, which is
particularly suited to a relativistic generalization, and we shall reformulate
it in a relativistically covariant way. Then we shall discuss the
group-theoretical underpinnings of relativistic fractional spin; this will
lead us to discuss the possibility of quantizing spin  without using
anticommuting variables or spinors. We shall thus introduce, in Sect.III,
the path integral
quantization of  spin, which will allow us to derive the fermion propagator
without introducing spinors. In Sect.IV this will be used to turn the
nonrelativistic
path integral of Sect.II into one which describes an arbitrary number of
(first-quantized)
relativistic point
particles with generic spin and statistics. In Sect.V we will finally discuss
some of the ideas and problems in the formulation of a consistent relativistic
(second-quantized)
field theory with fractional spin and statistics.

Our approach will be throughout of a rather
explicit nature. We will not aim at mathematical rigour, but rather at
introducing basic ideas and techniques. Also, we will not try to give a
comprehensive or fair review of the subject, which is rapidly growing; rather,
we will present a subjective perspective. The reader is referred to the
numerous good review papers in this subject both for a survey of different
approaches, and exhaustive references to the
original literature\ref\revs{ A general review
similar in spirit to the present lectures is
S.~Forte, { \it Rev. Mod. Phys.} {\bf 64}, 193 (1992). A general review
emphasizing especially the nonrelativistics quantum and statistical mechanics
is the book by A.~Lerda, ``Anyons'' (Springer, Berlin, 1992); the relation to
Chern-Simons theory is reviewed by R.~Iengo and K.~Lechner {\it Phys. Rep. }
{\bf 213}, 179 (1992). A collection of original papers is reprinted in
``Fractional Statistics and Anyon Superconductivity'', F.~Wilczek, ed.
(World Scientific, Singapore, 1990); the applications to high-$T_c$
superconductivity are reviewed  both in the introduction to this book and by
J.~D.~Lykken, J.~Sonnenschein and N.~Weiss, {\it Int. J. Mod. Phys.} {\bf A6},
5155 (1991). In the sequel of these lectures we shall not attempt to provide
complete references to the original literature, and we will limit ourselves
to citing
the sources on which our treatment relies.}.
The lectures are scattered with
a few exercises, which the reader interested in getting a working knowledge
of the
field is invited to work out.
\vfill
\eject
\newsec{\bf
FROM THE NONRELATIVISTIC TO THE RELATIVISTIC THEORY}

In two space dimensions angular momentum
and statistics are not quantized\ref\nrel{ The presentation in this section is
essentially based on the seminal paper by
J.~M.~Leinaas and J.~Myrheim, {\it Nuovo Cim.} {\bf B 37}, 1 (1977),
which started the subject. The path-integral formulation was given by
Y.-S.~Wu,  {\it Phys. Rev. Lett.} {\bf 52}, 2103 (1984).}.
 This is an
immediate consequence of the symmetry structure of the wave function of a
two-dimensional system. The wave function carries a representation (reducible,
in general) of the rotation group,
 which in $d$ dimensions is the group O($d$).
If $d>2$, then O($d$) is doubly connected (\ie, $\pi_1[\hbox{O($d$
)}]=\Z_2$); its
universal cover is the group Spin($d$), which, in the usual $d=3$ case,
is isomorphic
to SU(2). But when
$d=2$, the rotation group is O(2), which, being isomorphic to the circle $S^1$
is infinitely connected ($\pi_1[O(2)]=\Z$); its universal
cover is the real line $\R$.
It follows that whereas in more than two dimension
the wave function can carry either a simple-valued or a double valued
representation of the rotation group,
in two dimensions it may carry an arbitrarily multivalued one.
This means that upon rotation by $2\pi$ the wave function may
acquire an arbitrary phase;
the multivaluedness
of the representation is classified by the value of this phase:
\eqn\rot
{R^{2\pi}\psi=e^{2\pi i \ell} \psi,}
\ie, by the parameter $\ell$.
Because rotations are generated by the orbital angular momentum
operator, the allowed values of $\ell$
provide the spectrum of allowed values of
the angular momentum. It follows that in more than two dimensions the
angular momentum can
only be integer or half-integer, whereas in two dimensions it can be generic.
We  are intentionally speaking of (orbital)
angular momentum rather than spin here; the
distinction shall become clear in a relativistic framework, in Sect.IV.

Also, the wave function for a system of $n$ particles carries a one-dimensional
unitary representation of $\pi_1({\cal C})$,
the fundamental group of the (quantum-mechanical) configuration space
${\cal C}$.  This representation is characterized by the phase
which the wave function acquires upon the interchange of any two particles,
which is universal if the particles are identical
if we denote by $q_i$ the set of quantum numbers carried
by the $i$-th particle, then
\eqn\st
{\psi(q_1,\dots, q_i,\dots,q_j,\dots,q_n)=e^{2\pi i\sigma}\psi(
q_1,\dots, q_j,\dots,q_i,\dots,q_n).
}
The parameter $\sigma$ in Eq.\st\ is called the statistics of the particles.

For $n$ identical particles
in $d$ dimensions $\cal C$ is obtained by identifying all sets of $n$
$d$-component vectors (which can be thought of as the particles' positions)
after having removed its subset $D$ of all
 configurations where two or more particles coincide\foot{The exclusion of
 points where two particles coincide can be performed without loss of
 generality because such points are measure zero in the configuration space.
More precisely, it can be shown \ref\tarr{C.~Manuel and R.~Tarrach,
Barcelona preprint UB-ECM-PF 22/92 (1992);
{\it Phys.
Lett.} {\bf B 301}, 72 (1993).}
that under reasonable
assumptions of regularity of the wave function the contribution to the path
integral from intersecting paths vanishes. Eq.\st\ shows however that if
$\sigma\not=0$ then the wave function is either vanishing or singular when two
particles coincide, which may be viewed as the manifestation of the exclusion
principle for generic statistics. It is interesting to observe that if $\sigma$
is not zero but not half-integer the exclusion principle is somewhat weaker:
for example, point particles can have contact (\ie, delta-like) interactions
in such case, whereas fermions cannot.}:
\eqn\conf
{{\cal C}={{\R^{dn}-D}\over S_n},}
where $S_n$ is the set of permutations of $n$ vectors. If $d>2$, then
$\pi_1({\cal C})=S_n$, which admits only two one-dimensional representations,
the trivial one, and the alternating one; these correspond
respectively to integer
or half-integer values of the statistics $\sigma$ in Eq.\st.
If $d=2$, then $\pi_1({\cal C})=B_n$, the braid group, which
admits an infinity of representations, and the statistics can be
arbitrary\foot{In this case the value of the phase depends of course on
whether
the interchange is performed by a clockwise or counterclockwise rotation of the
two particles around their center of mass. Conventionally the statistics is
defined assuming that the interchange in Eq.\st\ is counterclockwise.}

Because the interchange
 in Eq.\st\ can be performed by means of a rotation by $\pi$ of the two
 particles, a rotation by $\pi$ generated by the total orbital
angular momentum exchanges all couples of particles. This establishes a
relation between the value of $\sigma$ and the allowed spectrum of values of
$\ell$:
\eqn\ssta
{\ell=k+\sigma n(n-1);\qquad k\in\Z.}
Notice that this is not (yet) a spin-statistics connection, because $\ell$ is
the orbital angular momentum,
being the eigenvalue of the generator $\hat l$ of space rotations (as distinct
from spin, which generates rotations in an internal space).
\bigskip

\subsec{\bf The path integral approach}

In the non-relativistic case a system of particles with generic spin and
statistics can be easily constructed starting from a theory of particles with
ordinary (say, bosonic) spin and statistics. In general, the quantum dynamics
is entirely described by $S$-matrix elements, \ie,
the transition amplitudes from an initial state $|i\rangle$ to a final
state $|f\rangle$:
\eqn\smat
{S_{fi}\equiv\langle\psi_f|\psi_i\rangle,}
which can be expressed in terms of the quantum propagator $K(q^\prime,q)
\equiv\langle q^\prime,t^\prime|q,t\rangle$, where $q$ denotes a point in
configuration space (such as, \eg, the set of positions of all particles):
\eqn\sprop
{\eqalign{S_{fi}=
&\langle \psi_f|q^\prime t^\prime\rangle\langle q^\prime t^\prime|
q t\rangle\langle q t | \psi_i\rangle\cr
=&\int\!dq\,dq^\prime\, \psi^*_f(q^\prime) K(q^\prime,q) \psi_i(q).\cr}}
The propagator is in turn given in terms of the Lagrangian $L$ of the system
by the Feynman path integral
\eqn\fpi
{K(q^\prime,t^\prime;q,t)=\int_{q(t)=q;\>q(t^\prime)=q^\prime}\!\!\!\!
\!\!\!\!\!\!\!\!Dq(t_0)\>e^{i\int_t^{t^\prime}\!dt_0\,L[q(t_0)]}.}

Now, given a theory with bosonic spin and statistics, described by a Lagrangian
$L_0$, a theory with generic statistics is obtained by adding to $L_0$
an interaction term $L_t$:
\eqn\newlag
{L=L_0+L_t}
where the interaction $L_t$ is given in terms of the position vectors $\vec
x_i$
of the $n$ particles by\foot
{In the sequel our notational conventions shall be as follows:
latin indices take the values 1,2, while greek indices run from 0 to 2;
$x^1$, $x^2$ are space coordinates and $x^0\equiv t$ is the time coordinate;
the three-dimensional Minkowski metric  is $(+,-,-)$;
the vector notation always denotes the (two) spatial components of vectors;
$\rho, \phi$ are polar coordinates on the space plane;
repeated indices are summed over; $\epsilon^{ab}$ and
$\epsilon^{\mu\nu\rho}$ are, respectively, the two- and
three-dimensional  completely
antisymmetric tensors, with the convention
$\epsilon^{12}=\epsilon^{012}=1$; the exterior  product of two vectors is
defined as $\vec v\times \vec w$=$\epsilon^{ab}v^aw^b$ (notice that it is
a scalar).}
\eqn\ltop
{\eqalign{&L_t=-s\sum_{i\not=j}{d\over dt} \Theta(\vec x_i-\vec x_j)\cr
&\quad
\Theta[\vec x]=\tan^{-1}\left(x^2\over x^1\right).\cr}}
The function $\Theta(\vec x)$ is just the polar angle of the vector $\vec x$;
it is defined as a multivalued function on the punctured
plane $\R^2_p\equiv\R^2-\{0\}$; it is single-valued on its universal
cover $\widetilde{\R^2_P}$,
which is the
Riemann surface of the complex logarithm.
The choice of branch can be fixed
defining
\eqn\br
{\Theta(q)=\int_{q_0}^q
\!dq^\prime\,{d\over dq^\prime}\Theta(q^\prime),}
where $q\in \R^2_P$ is a point in the punctured plane spanned by
$\vec x_i-\vec x_j$ for all $i,j$, and the integration runs along a path
which joins a fiducial
reference point $q_0\in \R^2_P$ to the point $q$ at which $\Theta$ is
evaluated.
It is essential that the multivalued definition of the angle be taken
(\ie, if $\vec x$ is rotated by $2\pi$ then $\Theta$ also changes by $2\pi$)
even though the choice of branch (\ie, the choice of $q_0$) is immaterial.
\topinsert
\vskip 7.5truecm
\figcaptr{1:}{ Linking number of particles trajectories (solid lines). The
linking number is defined by joining the endpoints
to infinity along a fixed direction (dashed lines). a) $l=-1$; b) $l=0$;
c) $l=1$.
\smallskip}
\endinsert

$L_t$ is called a topological
Lagrangian because, being a total derivative, it leads to a  contribution to
the action (\ie\ to the path integral) which does not
depend on the details of the paths. Rather,  it   depends
on the endpoints, and, because of the multivaluedness
of the function $\Theta$, on
the topology of the paths. Indeed, $L_t$ is a sum
over all particle pairs of terms of the form
\eqn\lin
{l={1\over 2\pi}\int \! dt \,{d\over dt}\Theta(\vec x_i-\vec x_j),}
up to a coefficient of $-2\pi s$, where $l$ Eq.\lin\ is an expression
for the linking number of the curves $\vec x_i$, $\vec x_j$,
\ie, it is equal to the number of  times the two paths link (see Fig.1).
\foot{For open paths this may be defined by joining the endpoints to a point at
infinity along a fixed direction and in a fixed order.}

It is easy to check that indeed the
theory with Lagrangian $L$ \newlag\ describes generic spin and statistics:
the propagator of this theory is
\eqn\newprop
{\eqalign{K(q^\prime,t^\prime;q,t)&=\int_{q(t)=q;\>q(t^\prime)=q^\prime}\!\!\!\!
\!\!\!\!\!\!\!\!Dq(t_0)\>e^{i\int_t^{t^\prime}\!dt_0\,\left(L[q(t_0)]
-s\sum_{i\not=j} {d\over dt_0}\Theta[\vec x_i(t_0)-\vec x_j(t_0)]\right)}\cr
&=\sum_{n_{ij,\>(i\not=j)}\,=-\infty}^\infty
e^{-is\left(\sum_{i\not =j}\hat\Theta_{ij}(t^\prime)+2\pi n_{ij}\right)}
K_0^{(n)}(q^\prime,t^\prime; q,t)e^{is\sum_{i\not= j}\hat\Theta_{ij}
(t)},\cr}}
where for short $\Theta_{ij}=\Theta(\vec x_i-\vec x_i)$ and $\hat
\Theta\equiv[\Theta \mod \Z]$,
the sums over $n_{ij}$ correspond to contributions to
the path integral from paths that wind $n_{ij}$ times on the
configuration space,
and $K^{(n)}$ is the path integral \fpi, computed from the Lagrangian $L_0$,
but restricting the sum over paths in such a way that
for each set of values of  $n_{ij}$
only paths with the corresponding winding numbers are included.

Now, all the effects of the topological interaction can be absorbed in a
redefinition of the wave function: if we define
a new wave function
\eqn\nwf
{\psi_0(q,t)=e^{is\sum_{i\not=j}\Theta_{ij}(t)}\psi(q,t)}
then it is an obvious consequence of  Eq.\sprop\ that the same
$S$-matrix elements can be equivalently obtained
propagating the wave function $\psi$ with the propagator $K$,
Eq.\newprop, or propagating the new wave function $\nwf$ with the
usual propagator $K_0$. In other words, what the topological interaction
does is to lift the wave function from the configuration space
to its universal cover:
the wave function
$\psi_0$ \nwf\ at point $q$ carries a path joining $q_0$ to $q$;
roughly speaking, this
 allows it to ``remember'' along its
evolution the sheet of the Riemann surface on which it
should be evaluated.\foot{ More rigorously, the set of paths from $q_0$
to each point $q$ in configuration space defines a homotopy mesh, \ie, it
provides a unique prescription to close an open path, by joining its endpoints
to $q_0$. Then, an open path can be uniquely assigned to a homotopy class,
determined by its linking number (compare Fig.1). This, in turn,
determines the motion of the point on the Riemann surface, because the linking
number of the path is equal to the number of sheets the point has travelled in
the course of its motion.}
Hence, we constructed a theory which differs from the
starting one only by the boundary conditions satisfied by the wave function:
if we rotate by an angle $\alpha$ the wave function Eq.\nwf\ we get
\eqn\newrot
{R^\alpha\psi_0=e^{is\alpha n(n-1)} R^\alpha \psi.}
In particular, if $\alpha=2\pi$ comparing this with Eq.\rot\ shows that
if $\psi$ is left unchanged (\ie\ if we started with a theory of bosons) then
$\psi_0$ has angular momentum $j=sn(n-1)$. Thus, the topological
interaction has induced arbitrary angular momentum, and, due to the
relation Eq.\ssta,
fractional statistics as well.

Of course, the formulation with topological interaction and  ``conventional''
wave
functions $\psi$ is completely equivalent to that without interaction and
``twisted'' wave functions $\psi_0$, thus all the physical observables
are the same in the two cases. Consider in particular the angular momentum.
In the approach where there is no topological
interaction, but the wave function is $\psi_0$, which obeys the boundary
condition \newrot, the angular momentum operator has
the form $\hat l_0$ that it would have in a free
theory, but
the phase
$\Theta_{ij}$ provides an extra contribution to its spectrum.
In the approach where the wave function satisfies the usual boundary
condition, but there is a topological interaction,
the angular momentum
operator $\hat l$ receives a contribution from the topological
term, and it is
related to the angular momentum operator of a free theory by
\eqn\newam
{\hat  l=\hat l_0+s n(n-1),}
where the spectrum of $\hat l_0$ is the usual one (\ie\ the integers).
In both cases, the spectrum of angular momentum
is  given by Eq.\ssta\ with $\sigma=s$.
\bigskip
{\narrower\narrower \noindent {\bf Exercise:} Prove that the topological
Lagrangian
$L_t$ Eq.\ltop\ contributes to the canonical Noether angular momentum
$\hat l\equiv {dL\over d\dot q}\delta^R q$, where $\delta^R$ is the
variation of the Lagrangian $L$ upon
infinitesimal rotation. Show that the contribution shifts the angular  momentum
according to Eq.\newam.
\medskip}
\bigskip
\subsec{\bf Covariant formulation and Chern-Simons theory}

The path-integral approach discussed in the previous section is
non-covariant, in that it relies crucially on the non-covariant parametrization
of paths with time. It is also non-relativistic,
in that spin, \ie\ intrinsic angular
momentum for one-particle states is missing, even in the case of particles
which obey fermionic statistics. However, it may be derived from a fully
covariant formalism. This is accomplished in two subsequent
steps\ref\csref{The possibility of deriving fractional statistics from
the (covariant)
Chern-Simons interaction was first realized by
D.~P.~Arovas, R.~Schrieffer, F.~Wilczeck, and A.~Zee,
{\it Nucl. Phys.}
{\bf B 251}, 117 (1985). The presentation given here is oriented to
the relativistic and field-theoretic generalization, and based on
S.~Forte and T.~Jolic\oe ur, {\it Nucl. Phys}. {\bf B 350}, 589
(1991).}.
\bigskip
\noindent{\it From the Chern-Simons theory to the Hopf interaction}
\smallskip

We start with a theory of point particles with (say) bosonic statistics,
defined by the Lagrangian $L_0$. The point particle excitations are carried by
a current $j^\mu$ which may be written as a sum of Dirac deltas:
\eqn\curr
{\eqalign{j^\mu&=
\sum_{i=1}^n\left(1,{d\vec x_i\over dt}\right)
\delta^{(2)}\!\left(\vec x- \vec x_i\right)
\cr
&=\sum_{i=1}^n\int \!ds\,\delta^{(3)}\!\left(x- x_i\right)
{d x^\mu\over ds}.
\cr}}
Let us now construct a new theory, whose
Lagrangian is obtained by adding to the particle Lagrangian $L_0$
a coupling $L_c$ to an abelian gauge field whose dynamics is
provided by  the Lagrangian $L_f$:
\eqn\cscou
{\eqalign{&L=L_0+L_c+L_f\cr
&\quad L_c=e\sum_i\left(\dotx_i\cdot\vec A-A^0\right)\cr
&\quad L_f=-{1\over2s }
\Int 2 y \left(\vec A(\vec y)\times\dot{\vec A}(\vec y)+2A^0(\vec y) B(\vec y)
\right).\cr}}
The action $I$ associated to the Lagrangian \cscou\
may be written in covariant notation as
\eqn\csac
{\eqalign{&I=I_0+I_c+I_f\cr
&\quad I_c=\Int 3 x j^\mu(x)A_\mu(x)\cr
&\quad I_f=-{1\over 2s}
\Int 3 x \epsilon^{\mu\nu\rho}A_\mu(x)\partial_\nu A_\rho(x).\cr}}

The field action $I_f$ Eq.\csac\ is the Abelian version of the
Chern-Simons action.
Its peculiar properties
are due to the fact that the field
is coupled through the $\epsilon^{\mu\nu\rho}$ tensor, which is a generally
covariant object. It is often referred to as a topological action because of
its sensitivity to the global features of the gauge potential $A$.
For our purposes, however, it is enough to observe that the action
$I_f$ is quadratic in the field $A^\mu$. We can therefore compute
the path integral over the $A^\mu$ field exactly, \ie, we may determine the
effective action
\eqn\effac
{I_{\rm eff}[j]\equiv-i\ln\int\! {\cal D}A^\mu\, e^{i(I_c+I_f)}.}
The result is equal to the so-called Hopf action (the reason of the name will
be clarified in Sect.IV.1):
\eqn\hac
{I_{\rm eff}=I_H=\pi s\int\!d^3x\,d^3y\,j^\mu(x)K_{\mu\nu}(x,y)
j^\nu(y),}
where the bilocal kernel
\eqn\biker
{
K_{\mu\nu}(x,y)=-{1\over2\pi}\epsilon_{\mu\rho\nu} {(x-y)^\rho\over
|x-y|^3}}
is the inverse of the operator $\epsilon^{\mu\nu\rho}\partial_\nu$
when acting on the current $j^\nu$,
\ie, it satisfies
\eqn\invop
{\epsilon_{\mu\nu\rho}\partial_\nu K^{\rho\sigma}(x,y)=\delta_\mu{}^\sigma
\delta^{(3)}(x-y).}

\bigskip
{\narrower\narrower \noindent
{\bf Exercise}: a) Construct the Green function of the 2+1
dimensional Laplacian $G(x-y)$, which satisfies
$\partial_\mu\partial^\mu G(x-y)=\delta^{(3)}(x-y)$.\hfil\break
b) Prove Eq.\invop.
\medskip}
\bigskip

\noindent{\it From the Hopf interaction to the topological action}
\smallskip
Eq.\hac\ shows that the effect of coupling the point-particle current to the
Chern-Simons Lagrangian is to induce the current-current self-interaction
$I_H$. We show now that this, in the non-relativistic limit, leads back to
the topological interaction $L_t$, Eq.\ltop. First, we notice that because the
current \curr\ is a sum of deltas, the Hopf interaction reduces to a sum over
all pairs of particles:
\eqn\hopsum
{\eqalign{&I_H= s\sum_{i,j}I_{ij}\cr
&\quad
I_{ij}=-{1\over 2}\int \! dx_i^\mu\,dx_j^\nu \epsilon_{\mu\rho\nu}
{\left(x_i-x_j\right)^\rho\over|x_i-x_j|^3}.\cr}}

Then, we study the generic term $I_{ij}$ in the sum. This can be simplified
greatly by noticing that one can write
\eqn\dirmon
{{x^\mu\over|x|^3}=\epsilon^{\mu\alpha\beta}\partial_\alpha\tilde A_\beta(x).}
The function $\tilde A^\mu$ must be singular, because the l.h.s. of Eq.\dirmon\
may be written as a divergence. As a matter of fact, the l.h.s.
of Eq. (2.42) is the field of a Dirac magnetic monopole
and Eq.\dirmon\
defines $\tilde A$ as its potential,
which notoriously has a string of singularities (that can be put anywhere
by a choice of gauge). The geometrical reason for the appearance here
of the Dirac
monopole will be clarified in Sect.IV.
Anyway, for our purposes it is enough to pick a
particular form of $\tilde A$ that satisfies \dirmon; a convenient
one is
\eqn\monex
{\tilde A_0(t,\vec x)=0;\quad \tilde A_a(t,\vec x)=
-{\epsilon_{ab}x^b\over r(t-r)},}
where $r^2=|x|^2=t^2-x_1^2-x_2^2$.

Notice that at this step we are already
singling out time as special in that the string of singularities is put along
the time axis.
We also parametrize paths with time, and then, using the
expression \dirmon,\monex\ for the interaction kernel in the action
$I_{ij}$, Eq.\hopsum, we get
\eqn\finiij
{\eqalign{I_{ij}&=-{1\over 2}\int_0^T \!dt\int_0^T\!dt^\prime{dx_i^\mu(t)
\over dt}
\left(\partial_\mu\tilde A_\nu(x_i-x_j)-\partial_\nu\tilde A_\mu(x_i-x_j)
\right) {dx_j^\nu(t^\prime)\over dt^\prime}\cr
&=\int_0^T\!dt\, \epsilon^{ab}\left({dx_i^a\over dt}-{dx_j^a\over dt}\right)
{(x_i(t)-x_j(t))^b\over
|x_i(t)-x_j(t)|^2}+ I_g,\cr}}
where
\eqn\endp
{I_g=-{1\over 2}
\int_0^T\!dt\,\left(\tilde A_\mu(x_i(t)-x_j(T)){dx^\mu_i\over dt}-
\tilde A_\mu(x_i(0)-x_j(t)){dx^\mu_j\over dt}\right)+x_i\leftrightarrow x_j.}

\bigskip
{\narrower\narrower\noindent{\bf Exercise}: a) Provide
the intermediate steps in
Eq.\finiij\hfill\break
b) Prove that
\eqn\exer{\partial_a\Theta(\vec x)=-\epsilon^{ab}{x^b\over|x|^2}.}
\medskip}

Now we should distinguish two cases, either $i=j$ or $i\not=j$. If $i=j$ the
bilocal kernel in Eq.\hopsum\ looks singular when $t=t^\prime$.
However the last step
in Eq.\finiij\ shows that in fact when $i=j$ the entire integral vanishes.
This result can be arrived at in a more rigorous way by regulating the
divergence in the kernel. We will discuss this in Sect.IV,
where we shall see
that the vanishing of the self-intercation, even
though true in the nonrelativistic limit,
cannot hold in a relativistic treatment.
If instead $i\not=j$ we may use Eq.\exer\ to get
\eqn\resiij
{I_{ij}=-\int\!dt\,{d\over dt} \Theta(\vec x_i-\vec x_j)+I_g,}
which, up to the $I_g$ term, coincides with the topological action, \ie, with
the linking number Eq.\lin.
Notice that the assumption that $\Theta(\vec x)$ is a multivalued function is
implicitly made when using Eq.\exer, which  is correct
(as we will see in more
detail in Sect.V.2) only if a multivalued determination of $\Theta$ is used.

The terms $I_g$
vanish for closed paths; for open paths they are associated to a contribution
to the Lagrangian
which does not modify angular momentum and statistics (as it can
be explicitly verified by checking that it is rotationally invariant) and
need
not concern us here.
We have thus succeeded in reproducing the topological interaction $L_t$
by coupling the point-particle current to itself through a Chern-Simons field.
At this point we have gone as far as possible in making the nonrelativistic
theory of particles with fractional statistics and angular momentum look
covariant. In order to use this knowledge to construct relativistic quantum
mechanics with fractional spin we need a deeper understanding of the relevant
symmetry structure.  Before
we even try to construct such a theory, we must ask whether a relativistic
wave function may carry a multivalued
representation of rotations. But, just like a nonrelativistic wave function
carries
a representation of the rotation group, a relativistic one carries a
representation of the Lorentz group.
Hence we need to understand the
structure of the Lorentz group in 2+1 dimensions,
just like we did in the beginning of this section for the spatial rotation
group.

\bigskip
\medskip
\goodbreak
\noindent{\bf {\it Mathematical digression}: The Lorentz group in 2+1
dimensions}
\nobreak

\medskip
First we list some basic facts about the structure of the group\ref\wyb{See
\eg\
B.~G.~Wybourne,  ``Classical Groups for Physicists''
(Wiley, New York, 1974).}.
The generators in the fundamental representation are the $3\times3$
matrices
\eqn\gensf
{L^{(\mu\nu)}{}^\alpha{}_\beta=-i\left(g^{\mu\alpha}g^\nu{}_\beta
-g^{\nu\alpha}g^\mu{}_\beta\right).}
The operator ${1\over 2}(L^{(12)}-L^{(21)})\equiv R$
generates the compact rotation subgroup, while
the operators
${1\over 2}(L^{(0a)}-L^{(a0)})
\equiv B^a$ generate the non-compact boosts. The Lie algebra
is
\eqn\galba
{[B^a,R]=-i\epsilon^{ab}B^b,\quad[B^a,B^b]=i\epsilon^{ab}R,}
or, in covariant notation,
\eqn\galbb
{\quad[L^{(\mu\nu)},L^{(\rho\sigma)}]=i\left(g^{\mu\sigma}L^{(\nu\rho)}
+g^{\nu\rho}L^{(\mu\sigma)}-g^{\mu\rho}L^{(\nu\sigma)}
-g^{\nu\sigma}L^{(\mu\rho)}\right).}
This is the same as the Lie algebra of SL(2,$\R$):
\eqn\sltworal
{[X^0,X^+]=X^-;\quad [X^+,X^-]=-X^0;\quad[X^-,X^0]=-X^+
,}
 hence the two groups admit
the same universal cover.
SL(2,$\R$) is the group of matrices
\eqn\sltwor
{A=\left(\matrix{a&b\cr c&d\cr}\right),}
with real elements and such that $\det(A)=1$.

\bigskip
{\narrower\narrower \noindent{\bf Exercise}:a) Show that the condition
$\det(A)=1$ may be rewritten as the equation of a three-dimensional $(2,1)$
one-sheeted hyperboloid. This is the group manifold of the universal cover of
SO(2,1).
\hfill\break
b) Work out the correspondence between generators of SL(2,$\R$) and SO(2,1).
\hfill\break
c) Show that the elements of a rotation subgroup
of  SO(2,1) correspond to points on a ``neck'' of the hyperboloid.\medskip}

Because the group manifold of SL(2,$\R$) is a one-sheeted hyperboloid,
it follows that the Lorentz group in 2+1 dimensions is infinitely
connected:
$\pi_1 [SO(2)]=\pi_1 \hbox{[SL(2,$\R$ ) ]}=\Z $. Also, non-contractible paths
on the group manifold correspond to non-contractible paths in its rotation
subgroup. Hence,
multivalued representations of SO(2,1) correspond
to multivalued representations of rotations. Notice that if instead we
considered theories defined in Euclidean space-time the Lorentz group
would be SO(3). The group manifold
of SO(3) is notoriously doubly connected [the universal cover is SU(2)],
$\pi_1(SO(3))=\Z_2$. This implies that representations of SO(3)
can only be either single-valued or double-valued, that is, that spin
may be only either integer or half-integer: the Minkowski signature
of the metric is essential if we wish to consider fractional spin.

We can now look at the
irreducible representations (irreps)
of SO(2,1).
These are classified by the
eigenvalue of the Casimir operator ${\cal Q}=(B^1)^2+(B^1)^2-R^2$, and
obtained
diagonalizing the rotation generator $R$.

\bigskip
{\narrower\narrower\noindent{\bf Exercise}: Show
that the most general solution of the commutation relations
\galba\ has the form
\eqn\solcomm
{\eqalign{R\xi_m=&m\xi_m\cr
B^+\xi_m=&\sqrt{(d+m)(-d+m+1)}\xi_{m+1}\cr
B^-\xi_m=&\sqrt{(-d+m)(d+m+1)}\xi_{m-1},\cr}}
where the raising and lowering operators are defined in terms of
the boost generators as $B^{\pm}=B^1\pm iB^2$, and the eigenvalue of
${\cal Q}$ associated to a given irrep is $d(d-1)$. \medskip}

\noindent
All irreps are ladders of states of the form \solcomm.
These fall into three classes:

a)
If $d$ is integer or half-integer
then there is a $2d+1$-dimensional irrep, spanned by
$\xi_m$, $m= -|d|, -|d|+1,\dots, |d|$.
These are the analogue of the usual irreps of the rotation group in three
dimensions.

b) If $d$ is not integer or half integer there are two semi-infinite irreps,
bounded either from below or from above, and spanned respectively
by $m=d,d+1,d+2,\dots$ and $m=-d,-d-1,,-d-2,\dots$.

c) For every $0\leq d<1$ and every $d\not= j (\mod \Z)$ and
$d\not=- j (\mod \Z)$ there exists a doubly infinite irrep spanned
by $m=\dots,j-2,j-1,j,j+1,j+2,\dots$.

Unitary irreps  are obtained requiring
${B^\pm}^\dagger=B^\mp$ and $R^\dagger=R$.
The latter condition is satisfied only if
\eqn\uncoa
{\left(\xi_{m_1},\xi_{m_2}\right)=\alpha_{m_1}\delta_{m1,m_2}}
where $\alpha_{m_1}$ is a real positive constant. It then follows from
Eq.\solcomm\ that
\eqn\uncob
{\eqalign{\left(\xi_{m+1},B^+\xi_{m}\right)&=\alpha_{m+1}
\sqrt{(d+m)(-d+m+1)}\cr
\left(B^-\xi_{m+1},\xi_{m}\right)&=\alpha_{m}
\left(\sqrt{(d+m)(-d+m+1)}\right)^*.\cr}}
Thus unitary irreps are obtained
when the parameter $C_d=\sqrt{(d+m)(-d+m+1)}$ is real.

In the three above cases:

a) $C_d$  is purely imaginary, hence no representation is
unitary (except the trivial one $d=j=0$).

b) $C_d$ is real if and only if $d>0$; these irreps are
unitary.

c) $C_d$ is real either if $d={1\over 2}+ i\alpha$ (principal series of
representations) or if $d$ is a real number such that
${1\over2}-|j-{1\over2}|<d<{1\over2}+|j-{1\over2}|$
(supplementary series).

\bigskip
{\narrower\narrower \noindent{\bf Exercise}: prove
the conditions for unitarity a-c.
\medskip}

{}From this classifications it follows that there exist no finite-dimensional
unitary irreps.
Furthermore,
even if we are willing to give up unitarity (after all, the usual
spinor representation of the Lorentz group is not unitary)
finite-dimensional
representations are at most double-valued.
Hence if we try to generalize to arbitrary statistics the
usual route used in constructing theories of fermions, namely,
go to a wave function
which carries a true representation of the universal cover of the rotation and
Lorentz groups, we must introduce an infinite-dimensional wave function.
This is physically unpleasant, because if an infinite-component
wave function is to describe
a finite number of degrees of freedom, then it must be subject to an infinite
number of constraints.

However, this
rather unpalatable option can be avoided by taking advantage of an
alternative formulation of the spin dynamics which dispenses us from
considering wave functions defined  on the universal cover of the group.
To understand this, we must step back to the familiar case of fermions. We
will see that their dynamics can be formulated without ever introducing
wave functions which carry representations of the universal cover of the
rotation group, \ie, without using spinors. Rather, a formulation purely in
terms of bosonic variables is possible.
\vfill
\eject
\newsec{\bf SPIN WITHOUT SPINORS}

The classical and quantum dynamics of a spin-${1\over2}$ object is
traditionally formulated in terms of functions defined on the double
cover of the rotation group, \ie\ the group Spin($d$) in $d$
dimensions (in three dimensions this is SU(2)), \ie\
in terms of spinors. However, it also possible to formulate the
dynamics of spin in terms of phase-space variables.
This, upon quantization, leads to  a formulation of a spin path-integral in
terms of bosonic (as opposed to anticommuting) variables, and to a wave
function defined in phase-space, rather than on the group. When this formalism
is used to quantize spinning particles, one obtains phase-space wave
functions which carry the representations of the Poincar\'e group associated to
particle states without having to introduce wave functions defined on the
group.  We will first
present the phase-space approach to spin in the simple and familiar case of
a spin degree of freedom in three spatial dimensions; then we will discuss the
group theory which underlies the construction of the relativistic quantum
mechanics of point-particle states, \ie, the theory of the Poincar\'e group in
2+1 dimensions; and finally we will provide
a formulation of
the dynamics of 2+1 dimensional
fermions without using spinors, and show that it
is equivalent to the usual approach.
\bigskip

\subsec{\bf Path integrals for spin}

The formulation of spin in terms of phase-space variables is accomplished in a
Lagrangian framework, through the formulation of a spin
action\ref\niel{The path integral approach to spin was developed
by H.~B.~Nielsen and D.~Rohrlich  {\it Nucl. Phys.} {\bf B 299}, 471, (1988),
and further discussed by
K.~Johnson, {\it Ann. Phys. (NY)} {\bf 192}, 104 (1989).}.
 This leads
naturally to quantization in the path-integral approach. We study a single spin
degree of freedom in three dimensions.
Classically, this is defined as a system whose
only degrees of freedom are the components of the  angular momentum,
whose modulus is fixed. The configuration space is
the sphere $S^2$, which is convenient to view as the coset SO(3)/SO(2), and
to parametrize with spherical angles $\theta$, $\phi$ as\foot{
In this section, the vector notation denotes three-dimensional Euclidean
vectors, latin indices run from $1$ to $3$.}
\eqn\tgvec
{\vec e=\pmatrix{\sin \theta  \cos \phi \cr
\sin \theta\sin\phi \cr\cos \theta \cr}.}
The spin vector is then $ \vec J=s \vec e$.
Eventually, we shall be interested in the quantization of system which have as
a configuration space SO(2,1)/U(1), which, roughly speaking, is
the ``Wick rotation'' to Minkowski space of this, and is the one-sheeted (1,1)
two-dimensional hyperboloid spanned by
\eqn\tgvecmi
{\tilde {\vec e}=\pmatrix{\cosh \theta \cr\sinh \theta  \cos \phi \cr
\sinh \theta\sin\phi \cr}.}
To this purpose, we will use throughout a covariant notation, such that all
results carry over to the Minkowski case by replacing sines and cosines of
$\theta$ with their hyperbolic counterparts, and the coordinate $x^3$ with a
Minkowski coordinate $x^0$.
\bigskip
\noindent{\it An Ansatz for the spin action}
\smallskip
We first propose an Ansatz for the spin action
and explain its meaning, then we verify that it works both
classically and quantum-mechanically.

The action
\eqn\spinac
{I_s=\int\!dt\,{\cal L}(\theta,\phi)=s\int\!dt\,\cos\theta\dot\phi}
describes the spin dynamics both at the classical and quantum level. At the
classical level, this means that the action Eq.\spinac\ leads to the classical
spin canonical structure.

At the quantum level, this means that if the action
$I_s$ \spinac\ is used as a weight in the Feynman path integral, it leads to
spin quantization in the sense that it leads to the same $S$-matrix element
that one would obtain using the usual spin Hamiltonian. That is,
the $S$-matrix element are
\eqn\spinmea
{\langle f | i \rangle=\langle \phi_f  |e^{i\int H(t)\, dt}|\phi_i \rangle,}
where
\eqn\spinstv
{|\phi\rangle=|m\rangle\langle m|\phi\rangle;\quad \langle
m|\phi\rangle={e^{-im\phi}\over \sqrt{2\pi}} }
and $H(t)$ is a (generally time dependent) interaction Hamiltonian
such as, for example, the coupling to an
external magnetic field $H=s \vec J\cdot \vec B$;
they can be computed as
\eqn\spinmeb
{\langle f | i \rangle=\int_{\vec e(t_f)=\vec e(\phi_f);\>
\vec e(t_i)=\vec e(\phi_i)}\!\!\!\!\!\!\!\!\!\!\! \!\!\!\!\!\!\!
D \vec e \>\>\>\>e^{i\int\!dt\,{\cal
L}_s-V(\vec J)}}
where the boundary conditions can be imposed only on the value of $\phi$
(or of $\theta$), because quantum-mechanically $\phi$ and $\theta$ do not
commute (they determine different components of the angular momentum operator),
and, because the Lagrangian
${\cal L}_s$ is first-order in time derivatives, the potential $V$
coincides with the Hamiltonian.
More in general, path-integration with this weight gives the matrix elements of
functions of spin operators:
\eqn\melspin
{\langle f| F(\vec J)| i\rangle=
\int_{\vec e(t_f)=\vec e(\phi_f);\>
\vec e(t_i)=\vec e(\phi_i)}\!\!\!\!\!\!\!\!\!\!\!\!\!\!\!\!
D \vec e\>\>\> e^{i\int\!dt\,{\cal
L}_s-V(\vec J)} F(\vec J).}

The rationale for this Ansatz is clear if one considers the case of a closed
time evolution, \ie, one where the initial and final states coincide.
In such case, the spin action \spinac, evaluated along a
closed path $C$ (on the sphere $S^2$), equals
\eqn\solang
{\eqalign{I_s&=s\int_C\cos\theta\dot\phi dt=s\int_C\cos\theta d\phi \cr
&=s\int_S d\cos\theta d\phi=s\int_Sd\vec S\cdot \vec e=s\int_S\left({
\partial \vec e\over
\partial s} \times{\partial \vec e\over
\partial t}\right)\cdot \vec e,\cr} }
where $S$ is a surface on the sphere bound by $C$, we have used
Stokes' theorem, and in the last step we have introduced a parametrization of
the surface $S$ in terms of two parameters $s,t$. The last expression is
immediately recognized as the expression of
the solid angle subtended by the curve $C$. Thus the action $I_s$ Eq.\spinac\
is the analogue for a spin degree of freedom of the action for a free massive
particle: just like the latter, it is given by the simplest geometric invariant
of the trajectory. This is the arc-length for a particle, and the solid
angle  for a spin.

\bigskip
\noindent{\it The classical spin action and the Faddeev-Jackiw canonical
formalism}
\smallskip
We wish to check that the spin action $I_s$ \spinac\ leads to the canonical
structure of a classical spin degree of freedom, \ie, to the Poisson bracket
\eqn\amompb
{
\{J^i,J^j\}=\epsilon^{ijk}J^k.}
This can be done in a simple and elegant way through the Faddeev-Jackiw
formalism for the determination of the canonical structure (and quantization)
of systems with
Lagrangians which are first-order in time derivatives\ref\faja
{L.~Faddeev and R.~Jackiw, {\it Phys. Rev. Lett.}, {\bf 60}, 1692. A recent
review is given by R.~Jackiw, MIT preprint CTP-2215 (1993), to be published in
the proceedings of the workshop ``Constraint Theory and Quantization Methods'',
Montepulciano, Italy, 1993.}. The formalism applies whenever
the
Lagrangian can be written in the form
\eqn\fjlag
{L=f_i(x){d x_i\over dt} -V(x)}
where $x$ are phase-space variables and $f_i(x)$ are arbitrary functions.

The Euler-Lagrange equations for such a system are
\eqn\fjeleq
{{\partial V\over\partial x_i}=f_{ij} {d x_j\over dt}}
where
\eqn\fjfstr
{f_{ij}={\partial f_j\over \partial x_i}-{\partial f_i\over \partial x_j}.}
The Euler-Lagrange equations \fjeleq\ can be rewritten in canonical form
as
\eqn\fjelcan
{{d x^i\over dt}=\{x^j,x^i\}{\partial V\over \partial x^j}=\{V,x^i\}}
provided the Poisson bracket is defined as
\eqn\fjpb
{\{x^i,x^j\}=(f^{-1})^{ji}}
where $f^{-1}$ is the inverse of the matrix $f$ defined in Eq.\fjfstr.
Hence, in this formalism the variables $x$ are viewed as phase-space variables
(coordinates and momenta) with Poisson bracket given by Eq.\fjpb; in a
conventional treatment $x$ would be coordinates, whose conjugate momenta are
fixed by a constraint equation because the action is first-order. Of course,
after resolution of the constraints this would lead to the same results.

Let us now apply this to the case of the spin action \spinac, which can be
further rewritten by defining
\eqn\monop
{\vec e=\vec \nabla \times \tilde{\vec A}[\vec e].}
Here the potential $\tilde {\vec A}$ as a function of $\vec e$
is the same as the Dirac
monopole \dirmon,\monex\ discussed in the previous section. We won't however
need its explicit form. Using Eq.s~\monop,\solang\
the spin action becomes
\eqn\spinacb
{I_s=s\int_S\!d \vec S\,\cdot \vec e=s\int_S \!d\vec S\cdot\vec
\nabla \times \tilde{\vec  A}[\vec e]=
s\int_C\!dt\,{d\vec e\over dt}\cdot \tilde{\vec A}[\vec e].}
This has the form of the Eq.\fjlag, with
\eqn\fjdata
{\eqalign{f_i&=s\tilde A_i[\vec e];\cr
f_{ij}&=s\left(\partial_i\tilde A_j-\partial_j\tilde A_i\right)
=s\epsilon^{ijk}e^k\cr
f_{ij}^{-1}&={1\over s^2}f_{ij}.\cr}}
It follows immediately from Eq.\fjpb\ that
\eqn\epb
{\{e^i,e^j\}={1\over s}\epsilon^{ijk}e^k}
which, identifying $\vec J=s\vec e$,
coincides with the canonical angular momentum Poisson bracket
Eq.\amompb, which is what we set out to prove

\bigskip
\noindent{\it Geometrical formulation of the spin action}
\smallskip
In order to proceed to the quantization of the spin action, it is convenient to
introduce a little more formalism, and
rewrite the action
in a geometrically more transparent way, as an action defined on the space of
orbits upon group transformations. This has the advantage of leading directly
to
geometric quantization. To this purpose, we write the vector $\vec e(t)$
as the result of
acting on an arbitrary reference vector $\vec e_0$ with an SO(3) matrix
$\Lambda(t)$:
\eqn\eanlam
{\vec e(t)=\Lambda(t) \vec e_0.}
This fixes two out of the three Euler angles which parametrize $\Lambda$,
while leaving the angle corresponding to rotations around the $\vec e_0$ axis
undetermined. Eq.\eanlam\ also implies
\eqn\eander
{\dot {\vec e}(t)=\dot \Lambda(t) \vec e_0}
where the dot denotes total differentiation with respect to $t$.
The path traversed by the vector $e(t)$ on the sphere as a function of $t$
has thus been mapped to
a path traversed by the matrix $\Lambda(t)$ on manifold of
the coset SO(3)/SO(2) --- the space of orbits of $\vec e_0$ upon action of
$\Lambda$.

We can lift this to a path on the group manifold of SO(3) by fixing the
third Euler angle which determines $\Lambda$.
Because $\dot{\vec e}\cdot\vec e=0$, we may do it
by requiring that
\eqn\thirda
{\vec n(t)=\Lambda(t) \vec n_0}
where
\eqn\normv
{\vec n(t)={\dot {\vec e}(t)\over|\dot {\vec e}(t)|}.}
The vectors $\vec e(t)$ and $\vec n(t)$, together with
\eqn\binor
{\vec  b(t)=\vec e(t)\times \vec n(t)}
define an orthonormal frame, whose motion in space is generated by the action
of the matrix $\Lambda$.
We may exploit this to rewrite the spin action as an element of the Lie algebra
of SO(3), by defining further
\eqn\frenfr
{\eqalign{&\vec v^{(3)}=\vec e_0\cr
&\vec v^{(1)}=\vec n_0\cr
&\vec v^{(2)}=\vec b_0;\quad \vec b_0=\vec e_0\times \vec n_0,\cr}}
and choosing  the values
\eqn\reffam
{v^{(a)}_i=\delta^a_i,}
for the three vectors which form the reference frame.

\bigskip
{\narrower\narrower \noindent{\bf Exercise}:
Prove that:\foot{Notice that the results below hold true also in Minkowski
space with the obvious replacement of $\delta_{ij}$ with the Minkowski metric.}
a)
\eqn\lamanv
{\left(\Lambda^{-1}\dot\Lambda\right)^{ij}=
\vec v^{(i)}\cdot\dot{\vec v}^{(j)};
}
b) $\Lambda^{-1}\dot\Lambda$ is an element of the Lie algebra
(called the Maurer-Cartan form), \ie\ it can be
written as a linear combination of the generators
\eqn\lamalg
{\left(\Lambda^{-1}\dot\Lambda\right)_{ij}=\sum_{ab}
C_{ab}(M^{ab})_{ij},}
where
$(M^{ab})_{ij}=\left(\delta_i^a\delta_j^b-\delta_j^a\delta_{i}^b\right)
$ are the SO(3) generators in the fundamental representation and $C_{ab}$ are
three independent real constants;\hfill\break
c) the constants $C_{ab}$ are given by
\eqn\const
{C_{ij}={1\over4}\tr\left(M_{ij}\Lambda^{-1}\dot\Lambda\right)={1\over2}
\vec v^{(i)}\cdot\dot{\vec v}^{(j)};}
d) the spin action equals
\eqn\spinalt
{I_s=s\left(\tr\int\!dt\,{1\over2}
\left(\Lambda^{-1}\dot\Lambda M_{12}\right)+\hbox{integers}\right).}
{\bf Hint} to point d): prove first that $\int_S\left({
\partial \vec e\over
\partial s} \times{\partial \vec e\over
\partial t}\right)\cdot \vec e=\int \!dt\,\,\dot{\vec b}\cdot\vec
n+\hbox{integers}$. The meaning of the integers will be
discussed in Sect.IV.1.

\medskip}

Eq.\spinalt\ expresses the kinetic term in the spin action in terms of the
matrix $\Lambda$. A potential term, written in terms of $\vec J=s\vec e$,
can be expressed in terms of $\Lambda$ as well using the identity
\eqn\efrl
{e^i={1\over2}\epsilon^{ijk}\left(
\Lambda^{-1}{M_{12}\over2}\Lambda\right)_{jk}.}
The advantage of this formulation is that, once
the phase space
variables $\vec e$ are expressed
in terms of a dynamical group variable (\ie, the SO(3)
matrix $\Lambda$) the dynamics depends
only on the algebra of the group, hence, it does not depend on the
choice of a representation. Even though we can recover formulations in terms of
any group representation we please by choosing an explicit form of the
generators,
we are not forced to do so, which is ultimately
what we are
trying to accomplish.

As a simple exercise, let us see how the formulation in terms of spinors can be
recovered. To this purpose, choose the spinor representation
\eqn\spinrep
{M_{ij}=-i\epsilon^{ijk}\sigma_k}
where $\sigma_i$ are the usual Pauli matrices. Then
\eqn\lagspin
{\tr{1\over2}\left(\Lambda^{-1}\dot\Lambda M_{12}\right)=\tr\left(
\Lambda^{-1}\dot\Lambda{\sigma_3\over 2i}\right)=\left(
\Lambda^{-1}\dot\Lambda\left({\1+\sigma_3\over2i}\right)\right).}
But if we define two-component spinors
\eqn\refspin
{\psi_0=\pmatrix{1\cr0\cr};\quad \psi(t)=\Lambda(t)\psi_0,}
then the spin action (with $s={1\over2}$) reduces to
\eqn\spispi
{I_s={1\over2}\int\!{dt\over i}\,\psi^*(t){d\over dt}\psi(t),}
while the generic spin vector can be represented as
\eqn\spivec
{\vec e =\psi^*(t)\vec \sigma\psi(t).}

\bigskip
\noindent{\it Quantization of the spin action}
\smallskip
A detailed proof that indeed using the spin action $I_s$ in a Feynman path
integral leads to spin quantization according to Eq.s~\spinmeb,\melspin\
can be accomplished by explicit computation of the path integral\niel.
Rather than going through this rather elaborate procedure, we notice that, once
expressed in terms of the $\Lambda$ matrices, Eq.\eanlam, the quantization
of a spin degree of freedom is a particular case of the more general problem of
quantization of a system whose classical configuration space is the set of
orbits of a group $G$, in our case the rotation group.
It can be shown
that
for all such systems the axioms of quantum mechanics fix uniquely both
the structure of the Hilbert space, and the form of the action which
upon path integration yields quantization of the
system\ref\wieg{ P.~B.~Wiegmann, {\it Nucl. Phys.}
{\bf B 323}, 311 (1989).}.

Namely,
the Hilbert space of the quantized system is the
representation space of the universal cover $\tilde G$ of the given group $G$,
so that if $T(\tilde G)$ is a unitary representation of $\tilde G$
then the Hilbert space is spanned by the vectors $|\phi\rangle=T(g)|\phi_0
\rangle$
where $g$ are all elements $g\in\tilde G$.  Thus, all quantum evolutions of the
system can be viewed as trajectories traversed by $g$ in the
representation space, and the path-integral has the general form
\eqn\piwieg
{\langle f|i\rangle=\int\! Dg\, e^{iI_w[g]}}
with the given boundary conditions.
Furthermore, the axioms of quantum
mechanics, and in particular, the  principle of superposition of quantum
amplitudes $\langle f|i\rangle=
\langle f|f^\prime\rangle\langle f^\prime|i\rangle$ fixes uniquely
the kinetic term in the action in the path integral \piwieg:
\eqn\acwieg
{I_w[g]=\int\!dt\,\langle\phi_0|\left[
T(g^{-1}(t)){d\over i dt}T(g(t))-H(g(t))\right]|\phi_0\rangle,}
where $H(g)$ is a  Hamiltonian. Identifying $T(g)$
with the matrix $\Lambda$,
this is  recognized to coincide with the
form \spinalt,\spispi\ of the spin action.

We may understand this prescription in a
rough and ready way in the spin-$1\over 2$ case,
where it is easy to work out the path integral directly from the Hamiltonian
form, using the spinor formalism of Eq.\refspin-\spivec: a generic $S$-matrix
element has the form
\eqn\affl
{\langle f| i\rangle=\langle\psi_f|e^{i\int\! H \,dt}|\psi_i\rangle
=\prod_{j=1}^N\int\!d\Lambda_j
\langle\psi_{j+1}|e^{i\Delta t H(t_j) }|\psi_j\rangle,}
where the time interval has been sliced according to
$\Delta t={t_f-t_i\over N}$,
where $N$ eventually tends to infinity, $t_j=t_i+(j-1)\Delta t$,
and $\psi_j$ is given by Eq.\refspin\ with $\Lambda=\Lambda_j$. In the limit of
large $N$
\eqn\affcom
{\eqalign{\langle\psi_{j+1}|e^{i\Delta t H(t_i) }|\psi_j\rangle
&\approx
\langle\psi_{j+1}|\left(1+i\Delta t H(t_j)\right)|\psi_j\rangle
=1-{1\over2}\Delta t \psi^*{d\over dt}\psi +i\Delta t H(t_j)\cr
&\approx e^{i\left[\psi^*{d\over i dt}\psi- \Delta t H(t_i)\right]},}}
which gives the path integral \spinmeb\ with the form \acwieg,\spispi\
of the action.
This shows explicitly that the first-order action is obtained directly from
the time evolution of the state vectors, according to the prescription Eq.
\acwieg; because the action is already written
in terms of phase-space variables (coordinates and momenta) no integration over
momenta is required.

Finally, it is interesting to observe that
the form Eq.\spinalt\ of the action is that which leads
to the so-called
quantization of the coadjoint orbits\ref\coad{ See N.~Woodhouse, ``Geometric
Quantization'' (Oxford U.P., Oxford, 1980).}
of the group of which $\Lambda$ is an
element. In this formalism,
quantization is enforced by imposing the commutation relations which follow
from the Poisson brackets \epb; this leads to wave functions which are
characters of the given group. In the present case, these are
the  Wigner functions, \ie, the spin wave functions \spinmea-\spinstv.

The formalism for the quantization
of a three-dimensional spin degree of freedom
discussed so far reduces the problem to that of the quantization of the orbits
of a normalized vector upon SO(3) action. This suggests that by simply
Wick rotating SO(3) to SO(2,1) we may obtain quantization of the Lorentz group
in 2+1 dimension; furthermore, the quantization based on the spin action
Eq.\spinalt\ allows to abstract from the choice of a specific representation
and seems therefore to lend itself naturally to be generalized to the case of
arbitrary spin. However, if we wish to quantize spinning particles, rather than
a fixed spin degree of freedom, spin must be coupled to the translational
degrees of freedom. The way this is done is fixed by the representation
theory of the Poincar\'e group,
 since, according to Wigner\ref\wig{E.~P.~Wigner, {\it Ann. Math.} {\bf 40},
 149 (1939).}, one-particle states
are in one-to-one correspondence with Poincar\'e irreps.
We must therefore study this group and its representation theory.

\bigskip
\medskip
\goodbreak
\noindent{\bf {\it Mathematical digression}: The Poincar\'e group
and point particles }

{\bf in 2+1
dimensions}
\nobreak

\medskip
The Poincar\'e group
is the semidirect product of the Lorentz group and
the translation group; in $2+1$ dimensions it is the group ${\rm ISO}(2,1)=
T^3\otimes {\rm SO}(2,1)$\ref\bine{The representation theory of the
2+1-dimensional Poincar\'e group is worked out by
B.~Binegar, {\it J. Math. Phys.} {\bf 23}, 1511, (1982). For a general
treatment of
Wigner's method of induced representations see \eg\
A.~O.~Barut and R.~Raczka, ``Theory of Group Representations and
Applications'' (World Scientific, Singapore, 1986).}.
Its Lie algebra
is generated by the three generators $L^{(\mu\nu)}$
of the Lorentz group, and the three generators $P^\mu$ of translations
(physically interpreted as momentum operators), and
it is given by extending the Lorentz algebra (5.2) by the further
relations\foot{We revert henceforth to the notational conventions of Footnote
3.}
\eqn\poinal
{[P^\mu,P^\nu]=0,\quad[L^{(\mu\nu)},P^\rho]=i\left(P^\mu g^{\nu\rho}
-P^\nu g^{\mu\rho}\right).}
The Casimir operators are
\eqn\poincas
{P^2=P_\mu P^\mu u(p)\quad
W=\epsilon_{\mu\nu\rho}P^\mu M^{\nu\rho}, }
\ie\ the total momentum, and the Pauli-Lubanski scalar $W$
which generates rotations around the momentum axis
(and
is a vector
in the familiar 3+1 dimensional case).
The group is infinitely connected, its universal cover
$\widetilde{\rm ISO(2,1)}$ is obtained by taking the semidirect product of
translations with the universal cover $\widetilde{\rm SO(2,1)}$ of the Lorentz
group:
$\widetilde{\rm ISO(2,1)}=T^3\otimes \widetilde{\rm SO(2,1)}$.

The unitary irreps of $\widetilde{\rm ISO(2,1)}$ may be easily
classified and constructed
through Wigner's method of induced representations. According to this method,
all unitary irreps of the Poincar\'e  group (or its universal cover) are
induced by unitary irreps of the stability group of an orbit under Lorentz
action of a point in the space $\tilde N$  dual to a carrier space $N$
of an irrep of the translation group. This means that all Poincar\'e irreps are
constructed through the following procedure:\hfill\break
a) List all unitary irreps of $T^3$. These have all the form of momentum
eigenfunctions, \ie, plane waves
$e^{iv\cdot p}$ where $v\in T^3$. They are classified by the values of the
momentum eigenvalues, \ie, all vectors $p\in\tilde T$ which span the dual
space $\tilde T$.\hfill\break
b) List all the distinct orbits in $\tilde T$. These are all the distinct
sets of momentum values which can be obtained by acting on a reference momentum
vector with a generic Lorentz transformation. They are classified by all the
distinct eigenvalues $m^2$
of the total momentum $P^\mu P_\mu$.\hfill\break
c) Construct  all the distinct irreps of the stability subgroup of the
momentum vector, \ie, the subgroup of SO(2,1) which leaves that
vector invariant. These are generated the operator $W$
Eq.\poincas, hence classified by all its distinct eigenvalues $sm$.

Physically, $m$ and $s$ are interpreted as the mass and spin of the particle,
respectively, and each distinct irrep provides the wave function $u(p)$
of a
one-particle state with fixed mass and spin, which is thus an
eigenfunction of
the two Casimir operators:
\eqn\casei
{\eqalign{P_\mu P^\mu u(p)&=m^2 u(p)\cr
\epsilon_{\mu\nu\rho}P^\mu M^{\nu\rho} u(p)&=
msu(p).\cr}}
The transformation properties of
these wave functions are easy to construct explicitly following the above
procedure.
The transformation of $u(p)$ upon  translation
along $a^\mu$ ($a^\mu a_\mu=1$) is
\eqn\trantras
{e^{i\epsilon P\cdot a}u(p)=e^{i\epsilon p\cdot a}u(p).}

\bigskip
{\narrower\narrower \noindent{\bf Exercise}: a) Prove
that {\it if} when
$p_0=\pmatrix{m\cr0\cr0\cr}$,
the action of a rotation by $\theta$ on $u(p)$ is
given by a certain representation function $D_s[\theta]$
according to
 $e^{i\theta R} u(p_0)=D_s[\theta]u(p_0)$
{\it then} the action of a generic Lorentz transformation
$U(\Lambda)$ on a generic state $u(p)$ is
\eqn\indwig
{U(\Lambda) u(p)=D_s[\Gamma^{-1}(p)\Lambda\Gamma(\Lambda^{-1}p)]
U(\Lambda^{-1}p)}
where $\Gamma(p)$ is the Lorentz transformation which
takes $p_0$ to $p$: $\Gamma(p) p_0=p$.\hfill\break
b) Prove that for an infinitesimal rotation $R(\epsilon)$
and an infinitesimal boost $B(\epsilon \vec\theta)$ along $\vec\theta$
($|\vec\theta|=1$)
\eqn\comm
{\eqalign{\Gamma^{-1}(p)R(\epsilon)\Gamma(R^{-1}(\epsilon)p)&=R(\epsilon)\cr
\Gamma^{-1}(p)B(\epsilon)\Gamma(B^{-1}(\epsilon\vec \theta)p)&=R\left(
\epsilon{\epsilon^{ab}\theta_ap_b\over E+m}\right).\cr}}
}

The transformations
upon  infinitesimal rotations and boosts are determined by Wigner's
procedure, according to Eq.s~\indwig,\comm\ in terms of the representation
functions $D_s$ of the stability subgroup. Because this is just the abelian
rotation group U(1), $D_s(\theta)=e^{is\theta}$, and the transformations
upon infinitesimal rotations and boosts are
\eqn\transrb
{\eqalign{e^{i\epsilon  R} u(p)&=e^{is\epsilon}u(e^{-i\epsilon R}p)\cr
e^{i\epsilon \theta_aB^a} u(p)&=e^{is\epsilon
\left({\epsilon^{ab}\theta_ap_b\over E+m}\right)}u(e^{-i\epsilon  \theta_aB^a}
p)\cr}}
where $E$ denotes the time component of $p_\mu$ (the energy).
\bigskip

\subsec{\bf The relativistic spinning particle}

One-particle states carry, according to Wigner, irreps of the Poincar\'e group,
or generally its universal cover.
Poincar\'e representation theory tells us that such irreps correspond to
eigenstates of the two Casimir operators $P^\mu P_\mu$ and $W$; the eigenvalue
of the former is interpreted as the square mass whereas the eigenvalue of the
latter, which is the generator of rotations around the axis defined by the
particle's momentum, is the product of the particle's mass and spin.
This suggests that quantization of a spinning particle can be accomplished by
supplementing the quantization of the translational degrees of freedom of a
relativistic (massive) spinless particle with the further quantization of
the spin degree of freedom, which is just that corresponding to rotations
around the momentum axis. It is quite conceivable that this, in turn,
should  be
accomplished by the procedure which we described in Sect.III.2.
\bigskip
\noindent{\it An Ansatz for the spin action}
\smallskip
We posit an Ansatz for the action of a relativistic spinning particle, based on
Poincar\'e representation theory\ref\bala{This approach to
the quantization
of relativistic spinless and spinning particles was developed by
A.~P.~Balachandran, G.~Marmo, B.-S.~Skagerstam and A.~Stern,
``Gauge Theories and Fiber Bundles'' (Springer, Berlin, 1983). See also
A.~P.~Balachandran, G.~Marmo, B.-S.~Skagerstam and A.~Stern,  ``Classical
Topology and Quantum States'' (World Scientific, Singapore, 1991).}. We then
verify then that it leads to the
correct classical and quantum theory in the spin-$1\over2$ case.

The free massive spinning particle action is written as the sum
of the free spinless particle action $I_0=I_0[x,p]$ and the spin action $I_s$
Eq.\spinac. The spin vector is attached to the particle by requiring that
the constraint Eq.\casei\ be satisfied. Defining a generalized spin operator
\eqn\spinop
{S^\mu=\epsilon^{\mu\nu\rho}M_{\nu\rho},}
for a momentum eigenstate $u(p)$
with momentum
$p^\mu$
the constraint Eq.\casei\ takes the form
\eqn\spinei
{S^\mu e_\mu u(p)=s u(p),}
where we have defined a unit momentum vector
\eqn\unmom
{e^\mu\equiv {p^\mu\over m}.}

In other words, the constraint Eq.\casei\ just means that the spin and momentum
vectors are parallel, and the action takes the form
\eqn\rpac
{I=I_0[x,p]+I_s[e]-V[x,e],}
where $V[x,e]$ is a potential,
and $I_s$ is given by  Eq.\spinac, with $e$ expressed in terms of the momentum
according to Eq.\unmom.
The action \rpac\ can be written compactly in terms of the Lagrangian
\eqn\balalag
{L=p_\mu{dx^\mu\over dt} + s \tr\left( \Lambda^{-1}\dot \Lambda M_{12}
\right)}
with the constraints
\eqn\balacon
{\eqalign{&p^\mu p_\mu=m^2\cr
&\Lambda^\mu_\nu p_0^\nu=p^\mu;\quad p_0^\mu\equiv\pmatrix{m\cr0\cr0\cr}.\cr}}
Notice that using this constraint the Lagrangian may be expressed as
$L=L[x,\Lambda]$. The action \balalag\ is written in first-order form in order
to emphasize the analogy of the case of a spinning particle to that of a spin
degree of freedom discussed in Sect.III.1.
\bigskip
{\narrower\narrower \noindent{\bf Exercise}: prove that in the spinless case
$s=0$ the Lagrangian \balalag\ with Eq.\balacon\ is the first-order form
of
the standard Lagrangian for a
massive spinless particle
\eqn\laspinl
{L_0=m\sqrt{\dot x^2};}
 \ie, prove that after resolution of the constraints
these two Lagrangians coincide.
\medskip}

In the particular case of spin-$1\over2$ the spin action may be written in
according to Eq.\spispi\ in terms of a spinor $\psi$ \refspin,
while the constraint Eq.\spinei\
takes the form of a
Dirac equation:
\eqn\conspi
{\sigma^\mu p_\mu \psi =m \psi,}
where $\sigma^\mu$ are the Wick-rotated Pauli matrices, defined according to
Eq.\spinrep in terms of the Lorentz generators in the spinor representation,
and satisfy a 2+1 dimensional Clifford algebra, \ie, they
are 2+1 dimensional Dirac matrices.
\bigskip
\noindent{\it Classical theory}
\smallskip
It is very easy to verify that
the Balachandran Lagrangian Eq.\balalag-\balacon\ defines the classical
dynamics of a spinning particle in the free case. The classical equations of
motion for such a system are just the conservation laws for linear and
generalized angular momentum. The former
\eqn\consmom
{
\dot p^\mu=0}
follows trivially
by varying
the Lagrangian with respect to a generic variation of $x^\mu$.
Let us check the latter.

The most general variation of
$\Lambda$ is the infinitesimal Lorentz transformation
$\delta\Lambda=i\omega^{\mu\nu} M_{\mu\nu} \Lambda$, where $\omega^{\mu\nu}$
is an antisymmetric infinitesimal parameter matrix.
The variation of the Lagrangian is thus
\eqn\varlag
{\eqalign{&\delta L =-i
\tr \left(\omega^{\mu\nu}M_{\mu\nu} K\right)+{i\over2}\tr
\left(S{d\over dt} \omega^{\mu\nu}M_{\mu\nu}\right)\cr
&\quad K_{\mu\nu}=\left(\dot x_\mu p_\nu - x_\nu \dot p^\nu\right)\cr
&\quad S_{\mu\nu}=s\left(\Lambda^{-1}M_{12}\Lambda\right)_{\mu\nu},\cr}
}
where the trace refers to the matrix indices.
Hence the Euler-Lagrange
equation is
\eqn\consan
{{d\over dt}\left(x^\mu p^\nu-x^\nu p^\mu + S^{\mu\nu}\right)=0.}
This is just the conservation of the total (orbital and spin) angular momentum.
Eq.s \consmom\ and \consan\ show that indeed the classical equations of motion
of a free spinning particle follow from the Lagrangian \balalag-\balacon.
It may be further verified that by introducing minimal coupling to
an electromagnetic field according to the  replacement $p^\mu\to p^\mu-e A^\mu$
one obtains the correct coupling of a charged particle to an electromagnetic
field, hence the Michel-Bargmann-Telegdi equations of motion follow, and so
forth. We shall not pursue this further, and turn to the quantum theory.
\bigskip
\noindent{\it Quantization}
\smallskip
The Lagrangian \balalag,\balacon, which is written in first order form,
can be quantized along the lines discussed in the case of a
spin degree of freedom in Sect.III.1 \wieg.
Also, it may be shown \polya\ that the
path integral obtained from the action \rpac\
coincides with that of the so-called
Brink-di~Vecchia-Howe superparticle, which is equivalent to a spin-$1\over2$
particle. Finally, it may be verified \ref\jaku{T.~Jaroszewicz and
P.~S.~Kurzepa, {\it Ann. Phys. (NY)}, {\bf 210}, 255.}
that such path integral is equal to the
scaling limit of a sum over directed random walks (\ie, random walks with
Hausdorff dimension $d_H=1$), which is known to reproduce the Dirac propagator.
Rather than following any of these paths, we shall show explicitly (following
Ref.\polya) that this
path integral leads (in the free spin-$1\over2$ case)
to the known form of the Dirac propagator.

First, however, we need to rewrite the path integral for a free massive
spinless particle in a more convenient way. We start with the expression for
the Euclidean space propagator
\eqn\nspprop
{\langle x^\prime|x\rangle=\int_{x(0)=x;\>x(1)=x^\prime}\!\!\!\!\!\!\!\!
\!\!\!\!\!\!\!\!\!\!\!Dx(s)\>\>e^{-m\int_0^1\!ds\,\sqrt{\dot x^2}},}
obtained from the Lagrangian \laspinl, where $s$ is a covariant
parametrization of the paths $x(s)$, chosen so that the paths from
$x$ to $x^\prime$ are traversed as $s$ varies $0\le s\le 1$.
The path integral Eq.\nspprop\ can be rewritten by introducing a
function $g(s)\equiv \dot x^2$
which may be viewed as an induced metric along the curve with respect to the
parameter $s$, because it satisfies by construction
\eqn\inmet
{dx^2 = g(s)\, ds^2.}
We get
\eqn\propa
{\langle x^\prime|x\rangle=\int_{x(0)=x;\>x(1)=x^\prime}\!\!\!\!\!\!\!\!
\!\!\!\!\!\!\!\!\!\!\!Dx(s)Dg(s)\>\>\delta^{(\infty)}\! \left(\dot x^2-g\right)
e^{-m\int_0^1\!ds\,\sqrt{g}},}
where the the constraint Eq.\inmet\ along the
path is enforced for all $s$ by means of a  functional Dirac delta which we
have denoted by $\delta^{(\infty)}$.

Once written in terms of $g(s)$,
the path integral \propa\ is manifestly invariant
upon reparametrization of the paths: if we let $s\to f(s)$,
then, because of Eq.\inmet\ $g(s)\to g(f(s))[\dot f(s)]^2$. We may exploit this
invariance to choose a parametrization such that $g(s)=\hbox{const.}\equiv
L^2$.  The parameter $L$ is just the
length of the path, because
\eqn\pathl
{\int_0^1\!ds\, \sqrt{\dot x^2}
=\int_0^1\!ds\, \sqrt{g(s)}
=L.}
This may be viewed as a choice of gauge, which we can enforce by introducing
one more functional delta:
\eqn\propb
{\langle x^\prime|x\rangle=\int_0^\infty\! dL\,\int_{x(0)=x;\>x(1)=x^\prime}
\!\!\!\!
\!\!\!\!
\!\!\!\!\!\!\!\!\!\!\!Dx(s)Dg(s)\>\>\delta^{(\infty)}\!
\left(\dot x^2-g\right)
\delta^{(\infty)}\! \left(g-L^2\right)
e^{-mL},}
at the expense of an extra (ordinary) integration
over the path length.
The functional integration over $g$ has become trivial:
\eqn\propb
{\langle x^\prime|x\rangle=\int_0^\infty\! dL\,\int_{x(0)=x;\>x(1)=x^\prime}
\!\!\!\!
\!\!\!\!
\!\!\!\!\!\!\!\!\!\!\!Dx(s)\>\>\delta^{(\infty)} \left(\dot x^2-L^2\right)
e^{-mL}.}
We may trade the path integration over $x$ for a path integration over the
unit tangent vectors $e$ to the path
\eqn\tgvec
{e^\mu={\dot x^\mu\over |\dot x|}={\dot x^\mu\over L.}}
However, the constraint that the endpoints of
the path be at $x$ and $x^\prime$ is nonlocal when expressed
in terms of the tangent vectors $e$,
and must be enforced by an (ordinary three-dimensional)
Dirac delta. We get thus
\eqn\propc
{\eqalign{\langle x^\prime|x\rangle&=\int_0^\infty\! dL\,\int
De(s)\>e^{-mL}
\delta^{(\infty)}\! \left( e^2-1\right)
\delta^{(3)}({x^\mu}^\prime-x^\mu-\int_0^L\!ds\, e^\mu(s))\cr
&=\int\! dL\,d\vec p\int
De(s)\>e^{-mL}
\delta^{(\infty)}\! \left( e^2-1\right)
e^{i p\cdot \left(x^\prime-x-\int_0^L\!ds\, e(s)\right)},\cr
}}
which is the sought-for form of the spinless particle propagator.

We may now use the form Eq.\propc\ of the path integral
to construct that for spinning particles using the Lagrangian
\balalag. Because the path integral is already written in terms of the unit
momentum vector $e$, it is enough to add the spin action to the weight in the
sum over paths, while the constraints Eq.\balacon\ are automatically satisfied
if the spin vector is identified with $e$:
\eqn\spinpi
{\langle x^\prime|x\rangle
=\int\! d\vec p\,e^{i p\cdot \left(x^\prime-x\right)}\int
dL\,e^{-mL}\int De(s)e^{-i p\cdot \int_0^L\!ds\, e(s)}
e^{iI_s[e]}
\delta^{(\infty)}\! \left( e^2-1\right),
}
where $I_s[e]$ is the spin action Eq.\spinac, written in terms of the
parametrization Eq.\tgvecmi\ of the vector $e$.
Now, for a closed path
\eqn\spiavcl
{\int\!\!
D \vec e(s)\> e^{iI_s} \left[e_1^{\mu_1}(s_1)e_2^{\mu_2}(s_2)\dots
e_n^{\mu_n}(s_n)
\right]=\tr\left(\sigma_1^{\mu_1}\sigma_2^{\mu_2}\dots\sigma_n^{\mu_n}\right).}
\bigskip
{\narrower\narrower \noindent{\bf Exercise}: Prove Eq.\spiavcl. {\bf Hint}:
Prove first the cases $n=1$ and $n=2$ using the commutator Eq.\epb, then
proceed by induction.
\medskip}

In general, for an open path, we may use Eq.\melspin, which, being true for all
matrix elements, implies the equation at the operator level
\eqn\melsig
{\int D \vec e(s)\, e^{iI_s[e]} F(e^\mu)= F(\sigma^\mu)}
in the
spin-$1\over2$ case. Thus, using this result in Eq.\spinpi\ obtains
\eqn\dirpro
{\eqalign{\langle x^\prime|x\rangle
&=\int\! d\vec p\,e^{i p\cdot \left(x^\prime-x\right)}\int
dL\,e^{-mL}e^{-i p\cdot \sigma} \cr
&=\int\! d\vec p\,e^{i p\cdot \left(x^\prime-x\right)}{1\over\thru p+ m},\cr}
}
which is the Dirac propagator.

We have thus obtained the Dirac propagator starting with a formulation where
the spin degrees of freedom are expressed in terms of phase-space variables,
identified with the unit tangent vectors to the particle paths, and weighted
with the spin action discussed in the previous section. Even though in the
particular case of spin-$1\over 2$ this leads back to the usual formulation in
terms of spinors and Dirac matrices, our starting point, namely the action
\rpac\
seems to be valid for any value of the spin parameter $s$,
and does
not necessarily require the use of variables, like spinors, defined
on the cover of the gauge group. This is the formulation which we shall try to
generalize to the case of generic spin and statistics.

\vfill
\eject
\newsec{\bf POINT PARTICLES WITH GENERIC SPIN}

In Sect.II we have described an approach to the path integral for particles
with fractional spin and statistics
which seems to lend itself to a covariant formulation, in that it may be
derived from the covariant Chern-Simons action, Eq.\csac. In Sect.III we have
seen that spin-$1\over2$ path integrals may be formulated in a way which seems
to be amenable to generalization to the case of generic spin. We would now like
to merge these two approaches. What is missing from the treatment of Sect.III
is
the discussion of the arbitrarily multivalued representations of the Poincar\'e
group which are associated to particles with generic spin, as well as the
particle-particle interaction which is expected to lead to the generic values
of the statistics (and therefore, according to Eq.\ssta, of the orbital angular
momentum) when several particles with generic spin are present. On the other
hand, in the treatment of Sect.II whereas the statistics interaction follows
from a covariant coupling, what seems to be entirely missing is some
interaction which generates fractional spin in the case of a single particle:
indeed, in the nonrelativistic limit considered there, the Chern-Simons
coupling only leads to a coupling of each particle with each other particle,
and has no effect whatsoever for a single-particle system, while we know
from experience with the $s={1\over2} $ case,
that
the dynamical effects of spin are present
even for a single particle. Therefore, we shall first study the case of a
one-particle system, and
show that the spin action can be obtained from a covariant Chern-Simons
coupling. After pausing to describe the mathematical underpinnings of our
construction, we shall prove that indeed this lead to wave function which
carries the Poincar\'e irreps associated to generic values of spin and
statistics.
\bigskip

\subsec{\bf The Hopf action and the spinning particle}

Let us consider a system of a single particle, with action given by
Eq.\csac. In such case, the Chern-Simons coupling produces only one interaction
term of the form \hopsum, with $i=j$, \ie\
\eqn\selfcou
{
I_{ii}=
-{1\over 2}\int\!ds\,dt\, \epsilon_{\mu\nu\rho}{dx^\mu(s)
\over ds}
{\left(x(s)-x(t)\right)^\nu\over|x(s)-x(t)|^3}
{dx^\rho(t)\over dt},}
where the two integrations run along the same curve $x(s)$ traversed by
the particle, and $s,t$ are invariant parameters along the curve, for example
the arc-length $ds^2=dx^\mu dx^\nu g_{\mu\nu}$.
The bilocal kernel \biker\ in Eq.\selfcou\  is singular as $s\to t$.
Nevertheless,
expanding $x(s)$ in Taylor series in the vicinity of $s=t$ we
get
\eqn\taysing
{\epsilon_{\mu\nu\rho}\dot x^\mu(s)\dot x^\rho(t)
{\left(x(s)-x(t)\right)^\nu\over|x(s)-x(t)|^3}=-{1\over 6}|s-t|
\epsilon_{\mu\nu\rho}{\dot x^\mu(s)\ddot x^\nu(s)\ddot x\dot{}^\rho(s)
\over |\dot x(s)|^3}+O(|s-t|^2),}
where the dot denotes differentiation with respect to $s$.
This expression is $O(|s-t|)$ as $s\to t$, implying that the integrand
in Eq.\selfcou\
is regular, and actually vanishing when $s\to t$.

The particle self-coupling induced through the coupling to the Chern-Simons
term
is therefore perfectly well-defined, despite the singularity of the kernel
\biker. Whereas in the nonrelativistic limit the
self-coupling contribution was just set to zero,
we shall now show that in the relativistic case
it automatically produces the spin action discussed in the previous
section\ref\polfor{ The possibility of obtaining the spin action in the
spin-$1\over2$ case and the
fermion propagator from the self-coupling of a point-particle current through
the Chern-Simons term was discovered by A.~M.~ Polyakov,{\it
Mod. Phys. Lett.}
{\bf A 3}, 325 (1988). The theory in the case of generic spin and statistics
and with an arbitrary number of particles is worked out in the path-integral
approach in S.~Forte {\it Int. J. Mod. Phys.} {\bf A7}, 1025 (1992), whose
treatment we shall follow.}.
To this purpose, we must compute the integral Eq.\selfcou\ for a generic
space-time curve\ref\calu{The computation of  the integral Eq.\selfcou\
was first accomplished by  G.~Calugareanu,  {\it Rev. Math. Pure Appl.}
(Bucarest), {\bf 4}, 5 (1959),
who studied its application to knot theory. A
``physicist's'' discussion is given by
M.~D.~Frank-Kamenetski\u\i~ and A.~V.~Vologodski\u\i,
{\it Sov. Phys. Usp.} {\bf 24}, 679 (1981), which we follow here.}.
\bigskip

\noindent{\it The writhing number of a space-time curve}
\smallskip
Let us first consider, for simplicity, the case of a closed curve $x(s)$.
In order to  treat the singularity of the kernel Eq.\biker\ it is conveninent
to
introduce a ``framing'' of the curve $x(s)$,
\ie,  define a { \it new} curve
\eqn\fram
{x_\epsilon^\mu(s)=x^\mu(s)+\epsilon n^\mu(s),}
where $n^\mu$ satisfies
$n\cdot n=1$ and $n\cdot{dx\over ds}=0$ and $\epsilon\to0$ (see Fig.2).
Let us further define
\eqn\linfram
{I_\epsilon=-{1\over 2}
\int\!ds\,dt \epsilon_{\mu\nu\rho}{dx_\epsilon^\mu(s)
\over ds}
{\left(x_\epsilon(s)-x(t)\right)^\nu\over|x_\epsilon(s)-x(t)|^3}
{dx^\rho(t)\over dt}.}
This is just the integral $I_{ij}$ Eq.\hopsum, computed
for the two curves $x$ and $x_\epsilon$, which, according to Eq.\resiij,
is proportional to
their linking number $l$ Eq.\lin:
\eqn\hoplina
{I_\epsilon=-2\pi l(\epsilon),}
where $l(\epsilon)$ is an integer for closed paths (recall Fig.~1).
For $\epsilon$ sufficiently small, $l$ does not depend on $\epsilon$ (Fig.~2),
and we can define
\eqn\hoplinb
{l=-{1\over2\pi}\lim_{\epsilon\to0}I_\epsilon.}
\topinsert
\vskip 4truecm
\figcaptr{2:}{ Framing $x_\epsilon$ of a curve $x$.
\smallskip}
\endinsert

But  $l$ Eq.\hoplinb\ depends on the choice of framing, \ie, on the choice of
$n$ Eq.\fram. Clearly, this cannot be equal to $I_{ii}$ Eq.\selfcou,
\ie, $\lim_{\epsilon\to0}
I_\epsilon\not=I_{ii}$, because
$l$ Eq.\hoplinb\  is manifestly framing-dependent, while $I_{ii}$
is a well-defined integral with no reference to framing, hence
it must be   framing-independent.
This entails that
the integral and the limit in Eq.\linfram\ do not commute:
\eqn\idel
{\eqalign{I_\delta&\equiv-{1\over 2}\left[
\lim_{\epsilon\to0} \int\!ds\,dt-
 \int\!ds\,dt\lim_{\epsilon\to0}\right]
\epsilon_{\mu\nu\rho}{dx_\epsilon^\mu(s)
\over ds}
{\left(x_\epsilon(s)-x(t)\right)^\nu\over|x_\epsilon(s)-x(t)|^3}
{dx^\rho(t)\over dt}\cr
&\qquad =-2\pi l-I_{ii}
\not=0.\cr}}

We may determine $I_\delta$ by considering the integral which defines
$I_\epsilon$ according to Eq.\linfram, and
separating a small neighborhood of the point $s=t$ from its region of
integration:
\eqn\splita
{I_\epsilon=-{1\over 2}
\int_0^T\!ds\,
\left[\int_{s-\delta}^{s+\delta}+
\left(\int_{0}^{s-\delta}+\int_{s+\delta}^{T}\right)\right]\!dt\,
\epsilon_{\mu\nu\rho}{dx_\epsilon^\mu(s)
\over ds}
{\left(x_\epsilon(s)-x(t)\right)^\nu\over|x_\epsilon(s)-x(t)|^3}
{dx^\rho(t)\over dt},}
where eventually we shall let $\delta\to0$.
When $s\not=t$ the bilocal interaction kernel is alway regular, hence
the $\epsilon\to0$ limit commutes with the integration in the region with the
point $s=t$ excluded. It follows that
\eqn\splitb
{\eqalign{&-{1\over 2}\lim_{\epsilon\to0}
\int_0^T\!ds\,
\left[\int_{0}^{s-\delta}+\int_{s+\delta}^{T}\right]\!dt\,
\epsilon_{\mu\nu\rho}{dx_\epsilon^\mu(s)
\over ds}
{\left(x_\epsilon(s)-x(t)\right)^\nu\over|x_\epsilon(s)-x(t)|^3}
{dx^\rho(t)\over dt}=\cr
&\quad=-{1\over 2}
\int_0^T\!ds\,
\left[\int_{0}^{s-\delta}+\int_{s+\delta}^{T}\right]\!dt\,
\lim_{\epsilon\to0}\epsilon_{\mu\nu\rho}{dx_\epsilon^\mu(s)
\over ds}
{\left(x_\epsilon(s)-x(t)\right)^\nu\over|x_\epsilon(s)-x(t)|^3}
{dx^\rho(t)\over dt}\cr
&\quad=I_{ii}+O(\delta),\cr}}
where the last step follows from the vanishing of the integrand at $s=t$,
Eq.\taysing.
Comparing this with Eq.\idel\
shows that the noncommutativity, \ie\
the difference between $I_{ii}$ and the linking number $l$
comes entirely from the $s\approx t$
region, \ie
\eqn\idelcom
{I_\delta=-{1\over 2}\lim_{\delta\to0}\lim_{\epsilon\to0}
\int_0^T\!ds\,
\int_{s-\delta}^{s+\delta}\!dt\,
\epsilon_{\mu\nu\rho}{dx_\epsilon^\mu(s)
\over ds}
{\left(x_\epsilon(s)-x(t)\right)^\nu\over|x_\epsilon(s)-x(t)|^3}
{dx^\rho(t)\over dt}.}
\bigskip
{\narrower\narrower \noindent{\bf Exercise}: Prove that
\eqn\rescal
{I_\delta=
\int\!{ds\over 2\pi} \epsilon_{\mu\nu\rho}
e^\mu n^\nu {dn^\rho\over ds}\equiv2\pi\tau,}
by expanding the integrand in Taylor series around the point $s=t$ (compare
Eq.\taysing).
\medskip}

Using Eq.s~\idel,\idelcom,\rescal\ the
particle self-interaction is  found to be
\eqn\writh
{I_{ii}=2\pi\left(\tau-l\right).}
If we frame the curve using in Eq.\fram\
the principal normal, defined in terms
of the tangent $e$ as
\eqn\frnor
{n_p^\mu={\dot e^\mu\over |\dot e|};\quad e^\mu={\dot x\over |\dot x|},}
then $l$ is
called the {\it self-linking} number of the curve, while
\eqn\tors
{\tau=\int\!{ds\over 2\pi} b\cdot\dot n;\quad
b^\mu=\epsilon^{\mu\nu\rho}e_\nu n_\rho}
is the geometric torsion of the curve, while
$b$ is the binormal vector.
Then $I_{ii}$ Eq.\writh\ is called the writhing number of the given
curve.\foot{Strictly speaking this
terminology applies to the case of an Euclidean
metric, \ie, to the case of curves in three-dimensional space, rather than
$2+1$-dimensional space-time. There is, however, no obstacle to defining
linking, self-linking, etc. for a Minkowski metric, either by performing
the computations in Euclidean space and Wick-rotating the result
(supplementing appropriate factors of $i$), or by performing the computation
in Minkowski space directly.}

Eq.\writh, with the expressions \hoplinb\ and \rescal\ for the quantities $l$
and $\tau$ provides the desired expression of the interaction induced by
coupling a one-particle current to the Chern-Simons term.
Several remarks on this result are in order:\hfill\break
a) Even
though the interaction Eq.\writh\
has been obtained from a bilocal coupling of each point to each other
point along the particle trajectory, according to Eq.\writh\
it may be expressed as the integral along
the curve of a local function of the
curve and its derivative. \hfill\break
b) The result Eq.\writh\ is independent of the choice of framing Eq.\fram\
which
we have introduced in order to arrive at it. However its decomposition into the
two terms \hoplinb\ and \rescal\ is framing-dependent; the framing dependence
of
these two contributions cancel against each other.\hfill\break
c) Whereas $l$ Eq.\hoplinb\ is a topological quantity (\ie, it is invariant
upon
small deformations of the curve), $\tau$ Eq.\rescal\ is a metric quantity (\ie,
it varies continuously upon small variations of the curve), thus so is also the
interaction $I_{ii}$ Eq.\writh.\hfill\break
d) The self-linking number (\ie\ $l$ [Eq.\hoplinb] computed when the curve is
framed with the principal normal) has a geometric interpretation\ref\pohl{
W.~F.~Pohl, 1968, {\it J.~Math.~Mech.} {\bf 17}, 975 (1968).}
as the number
of intersection of the curve with the envelope of its normals. It is a measure
of the number of coils which the curve forms.

If the curve $x(s)$ is open, rather than closed, Eq.\writh\ is still true, with
$\tau$ given by Eq.\rescal, while $l$ receives a correction, as in the
nonrelativistic computation Eq.\resiij, which is present because for an open
path there is a certain ambiguity in the definition of the self-linking number,
just as there is one in the definition of the linking number. It may, however,
be set to zero by a choice of phase of the wave function and will be neglected
henceforth.
\bigskip

\noindent{\it The writhing number and the spin action}
\smallskip

We may now proceed to our final step, and show that the particle
self-interaction, Eq.\writh\ reproduces indeed the spin
action\polfor. To this purpose, we must go back to the formalism introduced in
Sect.III.1: we define a frame of three vectors $e^\mu(s)$,
$n^\mu(s)$ and $b^\mu(s)$,
and construct a matrix $\Lambda(s)$ which generates the
time evolution of this frame when acting on a reference frame, which we may
take as the configuration of the given frame at initial time $t=0$
(as in Eq.s~\eanlam,\thirda). The vector
$e^\mu$ is now the unit tangent to the curve, while $n^\mu$ is the framing
vector introduced in Eq.\fram. In the particular case in which the curve is
framed with the principal normal this is the so-called Frenet frame of the
curve. Of course, vectors are normalized with respect to the Minkowski metric,
thus $\Lambda$ is an SO(2,1) matrix.
For convenience, we also introduce the
labelling Eq.\frenfr\ of
the three vectors of the frame as $v_{\mu}^{(\nu)}$, where both
indices are raised and lowered with the Minkowski metric, and we make the
choice Eq.\reffam\ for the reference frame.
Using the Minkowski version of Eq.s~\lamanv-\const\
it then follows immediately that $\tau$ Eq.\tors\
is given by
\eqn\tauexp
{\tau=\int\!{dt\over2\pi}{1\over 2i}\tr\left(\Lambda^{-1}\dot \Lambda
R\right),}
where $R$ is the generator of the rotation subgroup of SO(2,1)
defined as in Eq.\galba.
With this expression for $\tau$, the self-interaction
is very close to the form Eq.\spinalt\ of the spin action.

In order to show the complete equivalence, we introduce an explicit
parametrization of the matrix $\Lambda$ with Euler angles:
\eqn\eulera
{\Lambda(s)=e^{i\phi(s) R} e^{i\theta(s) B_2} e^{i\psi(s) R},}
where $B_2$ is the generator of boosts along the $y$ axis (Eq.\galba).
With this parametrization,
the angles $\theta(s)$ and $\phi(s)$ parameterize the tangent
vector $e$ according to Eq.\tgvecmi,
while $\psi$ determines the direction of the vector $n$ in the
plane orthogonal to $e$.
It is easy to work out the form of $\tau$, $l$ and $I_{ii}$:
\eqn\parres
{\eqalign{
\tau&=\int\!{ds\over2\pi}\,
\left( \dot \phi \cosh \theta+\dot \psi\right)\cr
l&=\int\!{ds\over 2\pi}\,\dot\psi\cr
I_{ii}&=\int\!ds\,\dot \phi \cosh \theta.\cr}}

Eq.\parres\
shows manifestly the framing-independence of the coupling
$I_{ii}$;
it also shows that $I_{ii}$ coincides
with the Minkowski form of the spin action Eq.\spinac. Thus coupling the
current of a theory of bosonic point particles to the Chern-Simons term
in the one-particle sector of the theory induces an interaction term in the
Lagrangian which is identical to that which, if the coefficient of the
Chern-Simons coupling in Eq.\cscou\ is fixed with $s={1\over2}$ coincides with
that which leads to quantization of
spin-$1\over2$ particles.
It is interesting to observe that Eq.\parres\ also clarifies
the relationship between the form Eq.\spinac\ (in terms of the angles $\theta$,
$\phi$) and the alternate form Eq.\spinalt\ (in terms of the matrix $\Lambda$)
of the spin action: the former coincides with $I_{ii}$, while the latter is
written as the sum of $\tau$ and a framing correction. The framing correction
is
equal to
$-l$ and  is what was alluded to as ``integers'' in Eq.\spinalt.

Because in the present approach
there is nothing special about the value $s={1\over2}$,
and since we know that in the non-relativistic, many
particle case the Chern-Simons coupling produces automatically nontrivial
statistics, this suggests that this approach will
lead to physical states which
carry generic spin and statistics automatically, by just taking an arbitrary
number of particles and and arbitrary value for $s$. Before we show that this
is indeed the case we pause to study some of the mathematics which underlies
the peculiar features of the ubiquitous spin action.

\bigskip
\medskip
\goodbreak
\noindent{\bf {\it Mathematical digression}: The spin action, the
Hopf map, and the Dirac monopole}
\nobreak

\medskip
Whereas the expressions Eq.\parres\ of $\tau$, $l$ and $I_{ii}$ are true with
any choice of framing, they take forms which have a particularly simple
geometrical interpretation when specific choices of framing are made.
One such choice we
have already discussed, and corresponds to taking for $n$ the
canonical (Frenet) normal, so that $\tau $ is just the (Minkowski analytic
continuation of) the torsion, and $l$ the self-linking number. It is important
to notice that in such case the third Euler angle $\psi$ takes values
$-\infty\le\psi\le\infty$, in keeping with the interpretation of $l$ as a
(self)linking number: when $n$ goes one full loop around $e$, then $l$
increases by one unit; and
analogously the total torsion is an increasing function
along the curve. Recalling the discussion of the Lorentz group in Sect.II.2,
this means that the matrix $\Lambda$ Eq.\eulera\ is actually an element of the
universal cover of SO(2,1), since $\psi$ parametrizes the compact rotation
subgroup.

An alternative simple possibility is to choose $n$ as the vector such that
\eqn\natsec
{\psi=-\phi.}
With this choice
\eqn\parresa
{\eqalign{
\tau&=\int\!{ds\over2\pi}\,
\dot \phi\left(  \cosh \theta-1\right)\cr
l&=-\int\!{dt\over 2\pi}\,\dot\phi\cr}}
(while $I_{ii}$ is of course unchanged). Then both $\tau$ and $l$ have simple
interpretations in the case of a closed curve:
$\tau$ Eq.\parresa\ is an expression for the solid angle subtended by the path
traversed by $e$ (or rather, its Euclidean counterpart)\foot{This definition of
solid angle and that of Eq.\solang\ correspond to the two possibilities of
defining the solid angle of a curve on a sphere as that of the surface bound
by that curve and containing the north pole or the south pole, respectively.
Otherwise stated, with the definition Eq.\solang\ the solid angle of a small
closed
circle around the north pole is close to $2\pi$, whereas with the definition
Eq.\parresa\ it is close to $0$.}. The value of $l$ instead gives the
homotopy class of the path traversed by $e$ after
its manifold of definition has been punctured to
remove the point $\theta=0$. In the Euclidean case, for instance, this is just
the linking number of a path on the sphere from which the north pole has been
removed, \ie, the number of times the path loops around the north pole.

All these mathematical structures have a simple interpretation in terms of the
so-called Hopf fibration\ref\aifordub{The relationship
between the Hopf fibration, the Dirac monopole, and
spin quantization is discussed \eg\ by
I.~Aitchison, {\it Acta Phys. Pol.} {\bf B 18}, 207 (1987)
[reprinted in ``Geometric Phases in Physics'', A.~Shapere and F.~Wilczeck,
eds. (World Scientific, Singapore, 1989)]
The application to  relativistic
particles with fractional spin is discussed by Forte
Ref.\polfor. For general mathematical background see \eg\
B.~Dubrovin, S.~Novikov and  A.~Fomenko,
{\it Modern Geometry} (Springer, Berlin, 1984).
}.
This consists of viewing the sphere $S^3$ as a
fiber bundle with base space $S^2$ and fibre $S^1$.\foot{ As usual we refer
to the Euclidean case which is geometrically simple, although all results can
be formally extended to the Minkowski case.} In the Euler angle parametrization
$S^2$ is spanned by $e$ \tgvec, \ie, by the Euler angles $\theta$, $\phi$,
whereas the fiber is spanned by the third Euler angle $\psi$. A choice of
framing is just a section of this bundle: in particular Eq.\natsec\ is the
so-called natural section of the bundle.

The expression Eq.\tauexp\ for $\tau$ is recognized as the parallel
transport (holonomy) with respect to
the induced U(1) connection along the fibre:
the connection   is given by the rotation component of the Maurer-Cartan form
(recall Eq.s~\lamanv,\lamalg), \ie
\eqn\conn
{\hat A_\mu=\tr{1\over 2i}\left[\Lambda^{-1}
\partial_\mu\Lambda(t) R\right],}
and the holonomy is
\eqn\holo
{2\pi\tau=\int\!dx^\mu \hat A_\mu,}
where the integration runs along a path traversed by $\Lambda$ on the bundle
space.
This can be decomposed in the motion along the instantaneous fibre, given by
$2\pi l$, and the induced motion (holonomy) due to the motion on the base
space,
given by $I_{ii}$. The latter physically is
the Thomas precession due to the motion
of the frame \frenfr\ in space.

The connection $\hat A_\mu$ is well-known in physics as the Dirac monopole
potential; the possibility of expressing it with different sections of the
bundle corresponds to the possibility of  choosing
gauge inequivalent potentials, because in that
application the fibre degree of freedom is viewed as a gauge degree of freedom.
As is well-known, this potential has in general a singularity,
which has a different location in different gauges. This is seen explicitly in
the various expressions
for $\hat A_\mu$ which we have given so far: for example
Eq.\parresa\ corresponds to choosing in Eq.\holo
\eqn\monopsp
{\hat A_\mu[e]=\left(0,-{\epsilon_{ab} e^b\over \left(e^0-1
\right)}\right),}
which is singular at the north pole, and is the same form of the monopole
potential which was used in Eq.\monex (notice however that in Eq.\monex\ the
monopole was in the space of positions, here it is in the space spanned by
tangent vectors). The choice of framing $\psi=0$, which gives $I_{ii}=\tau$,
hence
\eqn\holopr
{I_{ii}=\int\!dx^\mu \hat A^\prime_\mu}
corresponds to the choice
\eqn\monopsppr
{\hat A_\mu^\prime[e]=\left(0,-{\epsilon_{ab}e^0 e^b\over \left((e^0)^2-1
\right)}\right),}
which has singularities both at the north and south pole, and so forth.

The need for a singularity in the potential may be understood
by observing that
the integral of the connection \conn\ along a closed loop $C$ which bounds a
surface $S$ is
\eqn\fchern
{\eqalign{\oint \! dt\, {de\over dt}\cdot \tilde A[e]
&=\int_S d S^\mu\epsilon_\mu{}^{\nu
\rho}\partial_\nu\tilde A_\rho^\prime[e]\cr
&=\int_S d S^\mu\Omega_\mu,\cr}}
where $\Omega_\mu$ is the first Chern class of the Hopf bundle (the field of
the monopole). Because the Hopf bundle is nontrivial, the Chern class is closed
but not exact, hence the potential (connection) is not globally well-defined.
The two expressions Eq.\parres\ and Eq.\parresa\ of the holonomy of the
connection \conn\ correspond to two different options to avoid this
singularity. In Eq.\parres\ the potential is formulated on the full bundle
space, rather than on the base space only. In Eq.\parresa\ the potential is
expressed on the base space only, but the bundle is trivialized globally
by puncturing: by removing a point from the base space (the north pole of
the sphere) the bundle is globally trivial.

In the Minkowski case the full ``bundle'' space is the group manifold
of SO(2,1) discussed in Sect.II.1, \ie\ a (2,1) one-sheeted hyperboloid; the
base space, spanned by $e$, is a (1,1) two-sheeted hyperboloid; and
the fibre, parametrized by $\psi$, is still a circle $S^1$. Whereas
in the Euclidean
version, the base is simply connected, the bundle is doubly connected, and the
fibre is infinitely connected, in the Minkowski case,
the bundle and the fibre are infinitely connected, while the base is simply
connected. In the latter case, the holonomy Eq.\holo, evaluated along a path
$P$ on the bundle
space, is (if $P$ goes
through the identity, \ie, the point $\theta=\phi=\psi=0$)
an expression for the winding number $w$ of $P$ over that space (which is
infinitely
connected), because it is equal to the total projected
motion along the circle which is the non-contractible neck of the space:
\eqn\lorcoc
{w=\int_P\!{dt\over 2\pi}\,w(t);\quad w(t)={1\over 2i}\tr\left(
\Lambda^{-1}\dot\Lambda R\right).}

Equipped with this geometric knowledge we can now proceed to study the
path integral for particles interacting through the induced Hopf interaction
Eq.\hopsum\ in the case of generic values of the parameter $s$.
\bigskip

\subsec{\bf Path integral and multivalued relativistic wave functions}

We can proceed in a relativistic setting as we did
in Sext.II.2 for a nonrelativistic theory:
we start with a (now relativistic) theory of bosonic particles,
and we couple the covariant, conserved point particle current Eq.\curr\
to a Chern-Simons term according to Eq.\cscou, so
that effectively a bilocal current-current Hopf interaction of the form
\hac,\biker\ is generated. This, once the explicit form of the current
as a sum of $n$  Dirac deltas at the particles' locations
is used, is seen [Eq.\hopsum] to separate into $n$ particle
self-interaction terms $I_{ii}$, and $n(n-1)$ interactions of each particle
with each other particle. We shall first concentrate on the self-interaction,
by considering
the case of a one-particle  system, and show that for generic values of
the coupling parameter $s$ it indeed leads to fractional spin. Then we study
an $n$ particle system, see how generic statistics is also induced, and
finally discuss
the spin-statistics relation.

\bigskip
\noindent{\it The path-integral in the one-particle case}
\smallskip
For a one-particle system the induced Hopf interaction reduces to the single
term $I_{ii}$ which, as we have shown in the previous section Eq.\parres,
coincides with the spin action Eq.\spinac. Hence, the full
action coincides with the
relativistic particle action Eq.\rpac\ discussed in Sect.III.2, but now with
generic values of the spin parameter $s$. The path integral is thus
{\eqn\onepi
{\eqalign{K(x^\prime,t^\prime; x,t)=
&\int_{x(t)=x;\>x(t^\prime)=x^\prime}\sum_{n=-\infty}^\infty
Dx^{(n)}(t_0)\cr
&e^{-is
\left(\hat\psi(t^\prime)+2\pi n\right)}
\left[
e^{i\int_t^{t^\prime}\!\!dt_0\,\left\{L_0[x(t_0)]
+2\pi s \tau[e]\right\}}\right]
e^{is\hat\psi(t)},\cr}}
where $x$ is a point in 2+1 dimensional Minkowski space-time, $t$ is an
invariant parameter along the curve (such as the proper time), and
$\hat\psi\equiv\psi\mod \Z$.
If we use the canonical framing Eq.\frnor,\tors, the sum runs over
``self-linking classes'', \ie, paths are classified according to their
self-linking number, and for fixed $n$ only paths with self-linking equal to
$n$ are included in the path-integration.\foot{Notice
that the integration runs over all possible paths
from $x$ to $x^\prime$, including those which go backwards in time.
This means that the tangent vector $e$  may be space-like, hence
$\theta$ might be imaginary.}

The measure of integration over paths
of the $n$-th self-linking class $Dx^{(n)}$ is in practice rather
complicated. For practical purposes, it is more convenient to use the
framing Eq.\natsec; then, the spin action is entirely expressed according to
Eq.\parresa\ in terms of the tangent vector $e$, now defined on a punctured
hyperboloid, as discussed in the previous Sect.
It is thus convenient
to write the path-integral  with a separate
integration over the vector $e(t)$ along the path, and a
functional $\delta^{(\infty)}$ to enforce the constraint that $e$ be parallel
to the tangent
to the path at every point, as we did in the spin-$1\over2$ case in Sect.III.2:
\eqn\altonepi
{\eqalign{K(x^\prime,t^\prime; x,t)=
\int_{x(t)=x;\>x(t^\prime)=x^\prime}
D&x(t_0)e^{i\int_t^{t^\prime}\!dt_0\,L_0[x(t_0)]
}\cr
\times\int \sum_{n=-\infty}^\infty
De^{(n)}&(t_1)\delta^{(\infty)}\!\left({\dot x(t_1)\over|\dot x(t_1)|}
-e(t_1)\right)\cr
\times e&^{is
\left(\phi[e(t^\prime)]+2\pi n\right)}
\left[
e^{i2\pi s\int_t^{t^\prime}\!dt_1\,\tau[e(t_1)]}\right]
e^{-{is}\phi[e(t)]},\cr}}
where (using for simplicity the Euclidean nomenclature)
$e(t)$ traverses a curve on the unit sphere in the course of its
evolution, and $0\le\phi\le2\pi$ is an azimuthal angle on this sphere. The
$e$-integration in Eq.\altonepi\ is extended to all paths on this
sphere. Whereas $\tau$ is (for closed paths)
the solid angle subtended by
each path, the phase factor of
\eqn\self
{w={1\over 2\pi}\left[\phi(t^\prime)-\phi(t)\right]+n}
counts the total winding of each path about the axis through the poles,
and $n$ is just the homotopy class of the path on the sphere punctured
at the poles, which has fundamental group
$\pi_1\left(S^2-\{{\rm poles}\}\right)=\Z$.

The path-integral over $e$ in Eq.\altonepi\ is
akin  to that for the propagation of a charged
particle in the field of a Dirac monopole, with two important
differences: first, the monopole is in the space of tangent vectors,
rather than position space, and then, the topology of the
space is different. Whereas in the monopole case the singularity of the
action Eq.\parresa\ is treated by expressing it in terms of $\hat A$ according
to Eq.\holo, and then exploiting gauge invariance
to choose a form of the potential $\hat A$ which is always
free of singularities,
here the
singularity is treated by puncturing the sphere.
In other words, in the monopole case there
is no sum over $n$ in the path-integral Eq.\altonepi, and a single-valued
determination of $\tau$ is chosen by choice of gauge. These
two options correspond to different choices of space of quantization,
even though
the local canonical structure of the theory is the same.
A more careful treatment of the spin-$1\over 2$ case reveals
that the correct transition amplitudes, both for the Euclidean spin degree of
freedom \niel\ studied in Sect.III.1 [Eq.\melspin] and for the spinning
particle \jaku\ of
Sect.III.2 [Eq.\spinpi] are reproduced only if the multivalued prescription of
Eq.\altonepi\ is used, rather than the monopole prescription\foot{
The need for a multivalued phase may be understood as
the consequence of the fact that the coherence effects that yield the
desired quantization rules are effective only if one path-integrates over
a noncompact phase space.}. In the
Chern-Simons approach which we are pursuing this is an automatic consequence of
the computation.
We shall see shortly that this prescription is also mandatory if
we wish to obtain the multivalued Lorentz and Poincar\'e representations
associated to fractional spin.

We can now proceed to prove that in the case of generic $s$ the propagator
\altonepi\ defines the dynamics of a particle with fractional spin. We do this
by proceeding in analogy to what we did in Sect.II.1 in the nonrelativistic
case: we eliminate the ``spin'' interaction from the propagator by a suitable
redefinition of the wave function, and we show that the redefined wave function
carries the multivalued Poincar\'e irreps associated to generic spin.

\bigskip
\noindent{\it Elimination of the interaction}
\smallskip
It is clear that the nonrelativistic construction cannot be
reproduced literally, since the effect of the Hopf term is
not merely to endow the path integral with
the integral of a total derivative, as it
should necessarily be the case if the action Eq.\parres\
were purely topological.
Rather, the action Eq.\parres\
may be viewed as a Wess-Zumino term
\ie, as a total derivative in one dimension more, by rewriting $I_{ii}$
according to
Eq.s~\holopr,\fchern.

Then, the interaction can be eliminated, but at the expense of introducing a
wave function defined on a path, rather than on a point, \ie, defined
in one dimension
more. In particular, consider
a wave function $\psi(x,t)$ propagated by the path integral
Eq.\onepi\ according to Eq.\sprop, where now the configuration
space $\cal C$ is
that for a relativistic particle, \ie, $q$ is a point in 2+1 dimensional
Minkowski space-time, and $t$ is a covariant parameter along the curve
as in Eq.\onepi.  Then,
we define a new wave function $\psi_0$ which depends not only on
the point $x\in {\cal C}$, but also on a path $P_0$ that joins a reference
point $x_0$ to $x$:
\eqn\newwf
{\eqalign{&\psi_0(x)=e^{-is \Theta_{P_0}(x)}\psi(x);\cr
&\quad \Theta_{P_0}(x)={\int_{x_0}^x}_{P_0}\!dx^\prime\,
{de\over dx^\prime}
\cdot \hat A^\prime[e],\cr}}
where $\hat A^\prime$ is the Dirac monopole potential (in $e$-space) as given
by
Eq.\monopsppr, $x_0$ is a reference point in space-time, and $P_0$ is a path
that joins $x_0$ to $x$.

Because causality dictates that boundary conditions be imposed on
a space-like surface, the wave function must have support
in one such surface, hence the path $P_0$ must be contained in a space-like
surface, too.
\bigskip
{\narrower\narrower \noindent{\bf Exercise}: Prove that
if the path $P_0$ is planar, the phase $\Theta_{P_0}$ Eq.\newwf\
is invariant upon deformations of $P_0$.
\medskip}
\noindent
Without loss of generality, we may take the path $P_0$ to
be a straight line joining $x$ to spatial infinity along a space-like
plane.
The set of paths $P_0[x(t)]$ provides us with a mesh over space time
thereby allowing to reduce the computation of the spin
terms for an open path $P$ to the determination of the writhing number of the
closed  path $P_C$
which is obtained by joining the endpoints $x_i$, $x_f$ of
$P$ to $x_0$ through $P_0(x_i)$ and $P_0(x_f)$, respectively (recall
footnote 6).

The $S$ matrix elements computed for the wave functions $\psi_0$ and
$\psi$ are related in a simple way:
the former contains an extra weight in the sum over paths,
due to the transport of the phase $\Theta_{P_0}$ Eq.\newwf, which
equals
\eqn\extraph
{I_\Theta=-i\int\! dt\,\left(\langle\psi_0|{d\over dt}|\psi_0\rangle
-\langle\psi|{d\over dt}|\psi\rangle\right),}
where the integral runs along the given path.
Explicitly
\eqn\transp
{I_\Theta=s\int_P\!\!\! dt\,{d\over dt}\int_{P_0(t)}\cosh\theta\,d\phi
=s\int_S\!d\cosh\theta\,d\phi ,}
where $S$ is surface swept by the path $P_0(t)$ when $t$ runs
along the
path $P$. But this is  of course the same surface $S$ that appears
in Eq.\newwf, whereas the integrand is equal to that in Eq.\parres.
Hence, the $S$ matrix elements computed using in Eq.\sprop\
the wave function $\psi$ and
the propagator Eq.\onepi\
is identically equal to that computed by replacing
$\psi$ with $\psi_0$ and $K$ with $K_0$.

The Hopf interaction has thus been shown to amount to a phase redefinition of
the wave function. Because the Hopf interaction is not purely topological, in
that it contains the metric term $\tau$ Eq.\parres, the wave function has to be
lifted to a function defined on a path, rather than on a
point\ref\fromar{The fact that wave functions with fractional spin
must be localized on
a space-like line was proven by
J.~Fr\"ohlich and P.~A.~Marchetti,  Comm. Math. Phys.
{\bf 121}, 177 (1989) to be a consequence of the axioms of quantum mechanics}.
We should now
like to check that this is enough to endow the one-particle states of the
theory with the multivalued Lorentz and Poincar\'e  representations associated
to fractional spin. Before we do that, we would like to understand how this is
possible: in Sect.III we have shown how a path integral for spinning particles
can be constructed using phase-space variables, rather than variables on the
covering of the Lorentz group (\ie\ spinors); now, we would like to extend this
from path integrals to wave functions.

\bigskip
\noindent{\it Cocycles and Poincar\'e irreps}
\smallskip
The multivalued wave function Eq.\nwf\ introduced in order to construct the
nonrelativistic theory of particles with fractional statistics is a particular
example of a more general case\ref\jaccoc{Cocycles in quantum mechanics are
discussed by R.~Jackiw, in S.~Treiman, R.~Jackiw, C.~Zumino, and
E.~Witten
`` Current Algebra and Anomalies'' (World Scientific, Singapore, 1985). The
application to fractional statistics is treated by S.~Forte, Ref.\polfor\
and {\it Acta Phys. Pol.} {\bf B22}, 983 (1991).}.

Quite in general, consider a wave function $\psi(q)$ defined on a
certain configuration space $\cal C$, with a symmetry group $G$,
and such that $\psi$ is
single-valued when the group $G$ acts on the configuration space
(just like the rotation groups acts on the nonrelativistic configuration
space).
If the group $G$ is multiply connected, with universal cover $\tilde G$,
we may
construct a multivalued wave-function $\psi_0$ as
\eqn\multivalwf
{\psi_0(q)=e^{is\alpha_0(q)}\psi(q),}}
where $\alpha_0(q)$ is multivalued upon action
of the group $G$ on the configuration space.
In particular, if $g_0^n\in \tilde G$ is
in the $n$-th Riemann sheet of the group manifold of $\tilde G$, but
projects down to the identity of $G$,   we require that
\eqn\multivalph
{\alpha_0(q^{g_0^n})-\alpha_0(q)=n,}
where $q^g$ denotes the transform of point $q$ upon
action of the element $g$ of the group $G$.

If the action $U(g)$
of the group element $g$ on the
Hilbert
space spanned by wave functions is given by
\eqn\cantran
{U(g)\psi_0(q)=\psi_0(q^g),}
then the wave function $\psi_0$ Eq.\multivalwf\
upon group action transforms with an extra phase prefactor
(cocycle)
$\omega_1(q;g)$, according to
\eqn\cocycac
{U(g)\psi_0(q)=e^{i\omega_1(q;g)}\psi_0(q^g).}
The cocycle is given by
\eqn\cocycdef
{\omega_1(q;g)=s\left(
\alpha_0(q^g)-\alpha_0(q)\right)=s
\Delta^g\alpha_0.}

The phase Eq.\cocycdef\ is a 1-cocycle over the group, in that if we require
that the transformation law Eq.\cocycac\ preserves associativity of the group,
then $\omega_1(q;g)$
must satisfy
the 1-cocycle condition
\eqn\coccond
{\omega_1(q;g_2g_1)=\omega_1(q;g_1)+\omega_1(q^{g_1};g_2).}
Because of Eq.\cocycdef\ the condition Eq.\coccond\ is automatically satisfied.
A cocycle which may be expressed according to Eq.\cocycdef\ is
said to be trivial; nevertheless,
due to the multivaluedness Eq.\multivalph\ of $\alpha_0$ the triviality is only
local, \ie, it is not possible to eliminate the cocycle by a global
phase
redefinition of the wave function.

An explicit expression of
$\omega_1(q;g)$ can be given in terms of the winding number
density over the group G. The winding number density is a function
$w[g(t)]$ which, integrated along a non-contractible path over the group
manifold (which exists since by assumption $G$ is multiply connected) gives the
homotopy class of the path, \ie, an integer which identifies the class of
equivalence to which the path belongs.
Explicitly, if $P$ is a path over the group manifold of the $p$-th homotopy
class, which we may express as
a one-parameter smooth family of
elements of the group $g(t)$ parameterized
by $t$, then
\eqn\windef
{\oint_P\!{dt\over 2\pi}\,w[g(t)]=p.}
The cocycle is then constructed by choosing
a reference point $q_0$ in configuration space, and
it is given by integrating the winding number density along a path from $q_0$
to the given point $q$:
\eqn\cocwn
{\eqalign{&\omega_1(q;g)=s\int_{t_0}^{t_1}w[g(t)]\cr
&\quad q_0^{g(t_0)}\equiv\Lambda(t_0)q_0=q\cr
&\quad q_0^{g(t_1)}\equiv\Lambda(t_1)q_0=q^g.\cr}}

Because of Eq.\multivalph, the
wave function Eq.\multivalwf\  carries a multivalued
representation of $G$. The multivaluedness is
fixed by the value of the
parameter $s$. Hence, if we can find an expression for a function
$\alpha_0(q)$ such that the cocycle
Eq.\cocycdef\ computed from it is equal to the desired expression
Eq.\cocwn, then $\psi_0$ provides us with a wave function defined on
configuration space , but which (thanks to the phase prefactor) carries a
representation of the universal cover of the group.
In other words, the cocycle lifts the representation carried by the wave
function from the group to its cover. Thus
there is no need to define the wave function as
a function on the universal cover of the group, and the desired multivaluedness
is produced by the cocycle.

Hence, if the wave function $\psi_0$ upon Lorentz transformation
acquires a cocycle related according to Eq.\cocwn\ to
the winding number over the
Lorentz group (explicitly given by Eq.\lorcoc), then that wave function
carries a multivalued Lorentz representation, and generic spin.
The analogy with the nonrelativistic treatment, and the way both can be
understood within the framework which we just discussed, is summarized by the
following table.

\medskip
{\baselineskip 32pt
\tabskip=2em plus2em minus .5em
\halign to\hsize{\hfil#\hfil&\hfil#\hfil&\hfil#\hfil\cr
{\it general }&{\it nonrelativistic theory}&{\it relativistic theory}\cr
$G$& SO(2)&SO(2,1)\cr
$w(t)$ & $\Theta(\vec x_i-\vec x_j)$ &  ${1\over
2i}\tr\left(\Lambda^{-1}\dot\Lambda R\right)$\cr
$\alpha_0(q)$ & $ \Theta(q)=\int_{q_0}^q
\!dq^\prime\,{d\over dq^\prime}\Theta(q^\prime)$
& $\Theta_{P_0}(x)={\int_{x_0}^x}_{P_0}\!dx^\prime\,
{de\over dx^\prime}
\cdot \hat A[e] $ \cr}}
\medskip
\nobreak
\centerline{Tab.~1\quad The cocycle construction}
\goodbreak
\medskip
\bigskip
{\narrower\narrower \noindent{\bf Exercise}: a) Prove that
\eqn\transtet
{\Theta_{P_0}(\Lambda(g) x)=
\Theta_{P_0}( x)+
\int_{x}^{\Lambda(g) x} \!dx^\prime\,{de\over dx^\prime}
\cdot \hat A^\prime[e],}
where on the r.h.s. the integration  runs over a path of tangent vectors
obtained by acting on $e(x)$ with a path of matrices $\Lambda(t)$
that joins the unit of the group to
the given element $\Lambda(g)$ of SO(2,1).
\hfil\break
b) Prove that for a closed path in the space of $e$ vectors
\eqn\twococ{\int_{x}^{\Lambda(g) x} \!dx^\prime\,{de\over dx^\prime}
\cdot \hat A^\prime[e]=-\int_{t_0}^{t_1} {1\over 2i}\left(\Lambda^{-1}\dot
\Lambda R\right),}
where on the
r.h.s. of Eq.\twococ\ the integration runs on a path
on the group manifold which joins the identity to $\Lambda(g)$ (notice the
minus sign on the r.h.s.). {\bf Hint}: consider boosts and rotations separately
and use Eq.\comm.
\medskip}

It immediately follows from
Eq.s~\transtet\ and \twococ\ that the wave
function Eq.\newwf\ transforms with the Lorentz cocycle Eq.~\cocwn,\lorcoc,
as per the above
table. Notice that the cocycle is a function $\omega_1(e,g)$, \ie, it depends
on the group transformation $g$, and on the configuration-space point $q$
which is a tangent vector $e$, because the phase $\Theta_{P_0}$,
the monopole potential Eq.\conn, and the SO(2,1)
winding number Eq.\holo\ are defined as functions of $e$.
This entails that the wave function $\psi$ in Eq.\newwf\ ought to be defined as
a momentum eigenstate, with $e={p\over m}$.
In general, it should be Fourier decomposed in terms of
momentum eigenstates, each of which will carry a different phase
$\Theta_{P_0}$.

The Lorentz cocycle, when evaluated on a momentum
eigenstate, reproduces automatically the transformation law associated to
irreducible Poincar\'e representations according to Eq.\transrb.
\bigskip
{\narrower\narrower \noindent{\bf Exercise}: Prove that
\eqn\inftran
{\eqalign{\omega_1(e,\1+\epsilon R)&=s
\epsilon  \cr
\omega_1(e,\1+\epsilon\hat\theta^aB^a )&
=is\epsilon
\left({\epsilon_{ab}\hat\theta^ae^b\over 1+e^0}\right),
\cr}}
where $\omega_1$ is the cocycle given by Eq.\cocwn\ in terms of the SO(2,1)
winding number Eq.\holo. {\bf Hint}: use the explicit form Eq.\monopsp\ of the
monopole potential.
\medskip}

\noindent
In sum, the wave function $\psi_0$ Eq.\newwf\ carries the Poincar\'e irreps
associated to generic spin.

Because the formulation of the theory in terms of
the wave function $\psi_0$ and the propagator $K_0$ (without Chern-Simons-Hopf
interaction) is completely equivalent to that in terms of the propagator
Eq.\onepi\ and the conventional wave function $\psi$, the dynamics is seen to
admit a dual formulation, just as in the nonrelativistic case discussed in
Sect.II.1.
\bigskip
{\narrower\narrower \noindent{\bf Exercise}: a) Prove that the
Hopf interaction Eq.\hopsum\ with $I_{ii}$ given by Eq.\writh\
provides contributions
to the canonical conserved
Noether charges for angular momentum $R$ and boosts $B^a$
which have the form
\eqn\newnoe
{\eqalign{R_H=s\cr
{B_H}^a=s{\epsilon^{ab}e^b\over e_0+1}.\cr}}\hfil\break
b) Prove that if the operators $P_0^\mu$, $B^a_0$, $R_0$ satisfy the Poincar\'e
algebra Eq.s~\poinal,\galba, then also the operators
\eqn\newpoi
{\eqalign{P^\mu&=P^\mu_0\cr
R&=R_0+s\cr
B^a&=B^a_0+s{\epsilon^{ab}P^b\over P_0+m}\cr}}
satisfy the same algebra\ref\jacnai{R.~Jackiw and V.~P.~Nair, {\it Phys. Rev.}
{\bf D43}, 1933 (1991).}.
\medskip}
In the formulation in terms of $\psi_0$,
the spectrum of the angular momentum operator $J$ is shifted because of the
phase prefactor $\Theta_{P_0}$ in the wave function. In the formulation in
terms of $\psi$, it is the canonical angular momentum operator which is shifted
by the topological interaction  according to
\eqn\shrot
{R=R_0+R_H,}
where $R_0$ is the operator in the absence of Hopf interaction, and
$R_H$ is given by Eq.\newnoe. In the relativistic theory, a shift of the
angular momentum is consistent with the Poincar\'e algebra only if the boost
generators are shifted as well. Indeed, the spectrum of boosts is also shifted
according to
\eqn\shboo
{B=B^a_0+B^a_H,}
where again $B^a_H$ is given by Eq.\newnoe, and is due either to the
Hopf interaction which affects the operator, or to the phase prefactor
$\Theta_{P_0}$ which affects its spectrum. Because for momentum
eigenstates $e^\mu={p^\mu\over m}$ in terms of the momentum
eigenvalue $p^\mu$, the shift \shrot,\shboo\ can be viewed as a redefinition of
the Poincar\'e generators, which has the form Eq.\newpoi. Thus, the possibility
of introducing generic spin through a topological interaction is related to the
possibility of a redefinition of
the Poincar\'e generators which preserves the
algebra, but shifts the angular momentum spectrum.

\goodbreak
\bigskip
\noindent{\it Multiparticle states, spin and statistics}
\smallskip
\nobreak
We may proceed to study the general case of an $n$-particle system.
In this case, on top of
$n$ copies of the
spin action Eq.\parres\ the action
contains $n(n-1)$ particle-particle interaction terms.
These are actually the same which were discussed in the nonrelativistic case,
Eq.\resiij: indeed, the linking number of two space-time curves is covariant,
it is only the parametrization of paths with time introduced in Eq.\resiij\
which is not. Since, however, Eq.\resiij\ is clearly invariant with respect to
reparametrizations of the path (as it is manifest from its form
Eq.\br) it is enough to replace $t$
with any invariant parameter to obtain a covariant result.

In general, the explicit invariance will
be broken by the choice of boundary conditions;
however, Lorentz covariance is preserved: if, for
example, we impose on the path integral boundary conditions at fixed time
by requiring the initial and final states to be
$\langle\vec x;t|\psi_{i,f}\rangle=\psi_{i,f}(\vec x;t)$, then,
upon Lorentz transformation by $\Lambda$, the initial  and
final states become
$\langle \Lambda\vec x;\Lambda t|\psi_{i,f}\rangle=\psi_{i,f}(
\Lambda\vec x;\Lambda t)$.
The only effect of the linking-number terms is to endow
the path integral with multivalued phases $\Theta_{ij}$
Eq.\nwf\
which depend
on the endpoints of the path.
In a relativistic treatment, these
phases are defined as polar angles on the arbitrary
space-like plane on which boundary conditions at initial and final
times are imposed. Without further ado, we can give to all the results derived
in the nonrelativistic case in Sect.II.1 a Lorentz covariant
interpretation.
In particular, the propagator from $\psi_i(\vec x_1,\dots,
\vec x_n;t)$ to $\psi_f(\vec x_1^\prime,\dots,
\vec x_n^\prime;t^\prime)$ is
\eqn\npart
{\eqalign{&K(\vec x_1^\prime,\dots,
\vec x_n^\prime;t^\prime ;\vec x_1,\dots,
\vec x_n;t)=\cr
&\quad=\sum_{n_{ij},\>(i\not=j)\,=-\infty}^\infty
e^{-i\sigma\left(\sum_{i<j}\hat\Theta_{ij}(t^\prime)+2\pi n_{ij}\right)}\cr
&\qquad\qquad
\times\tilde K(\vec x_1^\prime,\dots,
\vec x_n^\prime;t^\prime ;\vec x_1,\dots,
\vec x_n;t)e^{i\sigma\sum_{i<j}\hat\Theta_{ij}(t)};\cr
&\quad\tilde K(\vec x_1^\prime,\dots,
\vec x_n^\prime;t^\prime ;\vec x_1,\dots,
\vec x_n;t)=\cr
&\quad\qquad
=\sum_{n_1=-\infty}^\infty\dots\sum_{n_n=-\infty}^\infty
\int
\left(\prod_{i=1}^n Dx_i(t_0)\right)\,e^{-is\left(\sum_{i=1}^n\hat\psi_i
(t^\prime)+2\pi n_i\right)}\cr
&\quad\qquad\qquad\times\left[
e^{i\int_t^{t^\prime}\!dt_0\,\left\{L_0(x_1(t_0),\dots,x_1(t_0))
+2\pi s\sum_{i=1}^n\tau_i[e_i]\right\}}
\right]\cr
&\quad\qquad\qquad\qquad\times e^{is\left(\sum_{i=1}^n\hat\psi_i
(t^\prime)\right)},\cr}}
where $\hat\psi$ and $\tau$ are as in the one-particle propagator Eq.\onepi,
$\hat \Theta_{ij}$ is as in Eq.\newprop,
and the parameter $\sigma=s$
has been introduced in order to ease the discussion of spin and
statistics.
In the non-relativistic limit the writhing number is ill-defined, since
the unit tangent to the curves in
three-space is always $e=(1,0,0)$, so that $\phi$ in Eq.\parres\ is
ill-defined.
We can then set $\phi=0$ conventionally, in which case
the nonrelativistic limit of the propagator \npart\ reproduces Eq.\newprop.

The extra multivaluedness introduced by the phases $\Theta_{ij}$ due to the
terms  particle-particle interaction
is the same as that which is present in the
nonrelativistic treatment and can be handled in the same way.
All the terms induced by the Hopf interaction in the propagator Eq.\npart\
may be
absorbed in a redefinition of the wave function which has the form Eq.\newwf\
for the self-interaction, and of Eq.\nwf\
case  for the particle-particle interaction.
The former phase is defined for momentum eigenstates, whereas the latter,
being a functional of the particle's positions is sharp for position
eigenstates. The redefinition of the wave function for, say, a position
eigenstate, takes thus the form
\eqn\newall
{\eqalign{&\psi_0(\vec x_1,\dots,\vec x_n;t)=\cr
&\qquad =e^{i\sigma\sum_{i=1}^n\sum_{j=1}^n
\Theta_{ij}(t)}
\int{m\over E_1} d^2k_1\dots
{m\over E_n}d^2k_n\, \cr
&\qquad\qquad\qquad\times e^{-is\sum_{i=1}^n\Theta_{P_0}(k_i)}
\langle k_1,\dots,k_n|
\psi(\vec x_1,\dots,\vec x_n;t)\rangle.\cr}}
Hence, both the multivalued phase which leads to fractional spin
$\Theta_{P_0}$ and that which leads to fractional statistics (and orbital
angular momentum) $\Theta_{ij}$ are generated by the same interaction,
expressed in terms of a  Dirac monopole potential (compare Eq.s~\newwf\ and
\dirmon, respectively). The  latter, however, is a function in the space  of
relatives positions of the particle, while the former is a function in the
space of tangent vectors to the particle trajectories (identified with
momenta).

Upon Lorentz transformation the wave function Eq.\newall\
acquires $n$ copies of the
cocycle Eq.\cocwn,\lorcoc, and
$n(n-1)$ phases due to the transformation of
$\Theta_{ij}$. These can be straightforwardly
shown to give again the same cocycle.
Thus,
upon Lorentz transformation the wave function
Eq.\newall\ acquires
$n+n(n-1)=n^2$ copies of the cocycle Eq.\cocwn,\lorcoc.
Upon spatial rotation, in
particular, the wave function acquires a phase which
shifts the total angular momentum spectrum $j_0$ of a theory without Hopf
interaction to
\eqn\amomsh
{j=j_0+[n+n(n-1)]s=j_0+n^2s.}

We can now finally discuss the spin-statistics relation.
It should be noticed that there are two, distinct relations between
statistics and angular momentum.
The first one is expressed by the fact that the operator $L_{x_ix_j}$
that generates rotations of the $i$-th and $j$-th particle about each
other is identified with their relative angular momentum operator.
This implies the relation Eq.\ssta\ between the statistics and the spectrum of
eigenvalues of the orbital angular momentum $\ell$.
This is a purely kinematical relation which is always true regardless
of the dynamics, and follows from the definitions of statistics and angular
momentum.
The second relation is a relation between the {\it spin}
angular momentum (which  may be measured, modulo integer, as the
total angular
momentum modulo integer of a {\it one}-particle wave function),
and the statistics of an $n$-particle wave function. Otherwise stated,
this is a relation between the values of the coefficients $\sigma$ and $s$
in the path integral and wave function Eq.~\npart,\newall.

The spin-statistics
theorem states that for Boson and Fermion fields $\sigma=s\quad\mod(\Z)$.
In the theory under investigation $\sigma=s$ and the
spin-statistics theorem is automatically satisfied. This is presumably
related to the fact that the theory can be derived as a particular limit of
the local field theory defined by the Cherns-Simons Lagrangian Eq.\cscou.
The orbital angular momentum $\ell$, total
spin $S$ and total angular momentum $j$
in our theory
are thus in general given  by
\eqn\spectra
{\eqalign{
\ell&=n(n-1)\sigma+k;\quad k\in\Z\cr
S&=ns\cr
j&=\ell+S,\cr}}
which, setting $\sigma=s$
 leads back to Eq.\amomsh.

This concludes our discussion of relativistic particles with fractional spin
and statistics. We have shown that the Hopf interaction induces both a particle
self-coupling, and a particle-particle coupling, which may be eliminated by a
phase-redefinition of the wave  function, provided the latter is localized on a
path. The redefined wave function carries a multivalued representation of the
Lorentz group. The multivaluedness is partly due to the fact that the wave
function carries multivalued representations of the Poincar\'e group due to
phases depending on each particle's momentum. It is
also partly due to the fact that
on the space-like plane on which boundary conditions to the quantum evolution
are defined the wave
function carries nontrivial representations of the braid group due to
phases depending on the relative position of each couple of particles.
The former leads to fractional spin, the latter to fractional statistics and
angular momentum. Because these phases are generated by the same interaction, a
spin-statistics relation holds. The wave function is defined in phase space,
and the multivaluedness associated to fractional spin and statistics
is generated by an interaction-induced phase cocycle which takes values on the
universal cover of configuration space, obtained acting on the configuration
space with the universal cover of the gauge group. This allows to obtain
multivalued representations of the Lorentz group without introducing a
wave function which carries a representation of the universal cover
of the group (as it is usually done for fermions), and which
would necessarily beinfinite-dimensional.\foot{A formulation in terms of
infinite-component wave functions with an infinite number of constraints is
however also possible\jacnai. Realistic dynamical calculations will presumably
have
to use a mix of the two approaches.}
\vfill
\eject

\newsec{\bf RELATIVISTIC FIELD THEORY}

Fractional spin and statistics in field theory may be introduced by considering
theories (like the O(3) model) which support localized topological solitons in
2+1 dimensions. Then, in the limit in which the separation of the solitons is
much larger than their size, fractional spin
and statistics may be  introduced by approximating  the solitons with point
particles and then proceeding as in the previous section. It is difficult to go
beyond this first simple step because of the lack of renormalizability of these
theories.

A perhaps more fundamental problem, however, is the construction of a
field theory whose elementary (point-like) excitations carry generic spin and
statistics. On the one hand, it is clear that most of the machinery which has
been introduced in the study of quantum mechanics with fractional statistics
will fail to work in field theory, because it relies heavily on the concepts
of particle trajectory, particle location, etc., which have no field-theoretic
analogue.
On the other hand, the construction of quantum mechanics with
fractional statistics described so far is based on the coupling of a conserved
current to itself by coupling it to a Chern-Simons term [Eq.\cscou] and
integrating the  Chern-Simons field out. This is a field-theoretic
construction, and one may hope that proceeding along the same lines when the
conserved current is a smooth field current, rather than a point-particle one,
will anyway lead to physical states with generic spin. A naive analysis
suggests that this is indeed the case, but encounters several difficulties.

Field theory with fractional spin and  statistics is still very much of an open
subject, and there exists no comprehensive
approach, even though attempts have
been made in several directions. Here we shall only sketch some of
the main problems,
and briefly described one possible avenue to solving
them, in order to give a feeling of the issues
which are
involved. We will first describe the simplest, canonical approach, display
the problems it runs into, and try to understand their origin. Then we will
discuss how these problems are resolved in a path-integral approach.
\bigskip

\subsec{\bf The Klein-Gordon-Chern-Simons theory}

We consider the simplest field-theoretic generalization of the theory of
particles coupled to Chern-Simons of Eq.\cscou: namely, a charged (complex)
scalar field
coupled to Chern-Simons.\foot{The field must be
complex
if we want to allow for generic statistics. Because
particles and antiparticles, which are generated by
complex-conjugate operators,
have equal and opposite spin and statistics, it follows that
a real
field is necessarily bosonic or fermionic.}
We consider its canonical
quantization and seek for the effects of the Chern-Simons coupling.\ref\semen{
This approach was developed by
G.~W.~Semenoff, {\it Phys. Rev. Lett.} {\bf 61}, 517 (1988),
{\bf 63}, 1026 (1989); see also
G.~W.~Semenoff and P.~Sodano, {\it Nucl. Phys.} {\bf B328}, 753 (1989).}

We start with the Lagrangian
\eqn\ftcs
{{\cal L}=\left(\partial_\mu+iA_\mu\right)\phi^*
\left(\partial_\mu-iA_\mu\right)\phi-m^2\phi^*\phi+{1\over
4\pi s}\epsilon^{\alpha\beta\gamma}A_\alpha\partial_\beta A_\gamma.}
The action associated to this Lagrangian can be written in the form Eq.\csac,
as the coupling to a Chern-Simons term of the field having action
\eqn\freeac
{I_0=\int
\!d^3x\,\left[\partial_\mu\phi^*\partial^\mu\phi-m^2\phi^*\phi\right]}
through  the conserved current
\eqn\concurr
{j^\mu(x)=i\left(\phi(x)\pi^\mu(x)-\phi^*(x)\pi_\mu^*(x)\right);\quad
\pi^*_\mu(x)=\left(\partial_\mu-iA_\mu\right)\phi(x).}

\bigskip
\noindent{\it Canonical quantization}
\smallskip

We want to quantize the theory canonically in the $A_0=0$ gauge.
To this purpose we must first discuss the constraints of the theory.
The equation of
motion for the $A_0$ field is a constraint (primary, first class),
because there is no time derivative of $A_0$ in the Lagrangian:
\eqn\primcon
{\pi(x)=0,}
where $\pi(x)$ is the momentum canonically conjugate to $\phi(x)$, \ie,
in terms of  $\pi_\mu(x)$ Eq.\concurr,
$\pi(x)=\pi_0(x)$.
Requiring this constraint to be preserved by the time evolution, \ie
requiring
$\{H,\pi(x)\}=0$,
 we get the
secondary constraint
\eqn\seccon
{j^0={1\over 2\pi s} \epsilon^{ij}\partial_iA_j,}
which is the Gauss law.
\bigskip
{\narrower\narrower \noindent{\bf Exercise}: Prove that
Eq.\seccon\ is solved by
\eqn\gausol
{A^i( \vec x,t)=-s
\int\! d^2y\,\epsilon^{ij} {(x-y)_j\over |\vec x-\vec y|^2}
j^0(\vec y,t).}
{\bf Hint}: Prove first that
\eqn\twolap
{\partial_i\partial_i
{1\over 2\pi}\ln |\vec x|=\delta^{(2)}(\vec x).}
\medskip}

The constraint Eq.\seccon\ determines
the nonvanishing components of the
Chern-Simons field $A^i(\vec x,t)$. Using Eq.\gausol\ and the property
Eq.\exer\
of the angle function $\Theta(\vec x)$ which we have repeatedly used in the
previous sections, we see that $A^i$ is given by
\eqn\puregau
{\eqalign{&A^i( \vec x,t)=
s\int\! d^2y\, {\partial\over\partial x^i}\Theta(
\vec x-\vec y)j^0(\vec y,t)=\partial_i sS(\vec x)\cr
&S(\vec x)=\int\!d^2y\,
\Theta( \vec x-\vec y)
j^0(\vec y),\cr}}
Thus, {\it if} the interchange of derivative and integral in the last step
of Eq.\puregau\ is
allowed, then
$A^i$ Eq.\puregau\ is a pure gauge, and may be removed by a
gauge transformation. This is however a highly nontrivial assumption, because
of the singular nature of the function $\Theta(\vec x)$, as we shall discuss
below.

Anyway, if we proceed naively, we conclude that the interaction with the gauge
field can be completely eliminated by defining the gauge-transformed fields
\eqn\gtran
{\eqalign{\phi_0(\vec x)&= e^{2is
S(\vec x)}\phi(\vec x)\cr
\pi_0(\vec x)&= e^{2is
S(\vec x)}\pi(\vec x),\cr}}
where $S(\vec x)$ is as in Eq.\puregau.

\bigskip
\noindent{\it Graded commutators}
\smallskip
Now we can impose canonical commutation relations
\eqn\cancomm
{\eqalign{[\phi(\vec x),\pi(\vec y)]
&=\delta^{(2)}(\vec x-\vec y)\cr
[\phi(\vec x),\phi(\vec y)]
&=[\pi(\vec x),\pi(\vec y)]=0,\cr}}
which imply in particular
\eqn\crop
{[j^0(\vec y),\phi^{\dag}(\vec x)]=
\delta^{(2)}(\vec x - \vec y)\phi^{\dag}(\vec x),}
\ie\ the field operator $\phi^*$ acts as a creation operator (it creates one
unit of charge). But then it follows that
\eqn\scomm
{\left[S(\vec x), \phi^{\dag}(\vec y)\right]=\Theta\left(\vec x-\vec y \right)
\phi^{\dag}(\vec y),}
hence the commutation relations satisfied by the gauge-transformed fields
$\phi_0$, $\pi_0$ Eq.\gtran\ differ from Eq.\cancomm, because of the extra
non-commutativity of the gauge function $S(\vec x)$.
\bigskip
{\narrower\narrower \noindent{\bf Exercise}: Prove that
\eqn\bacaha
{e^{-2is S(\vec x)}\phi_0^{\dag}(\vec y)e^{2is S(\vec x)}
=e^{-2is\Theta(\vec x-\vec y)}\phi_0^{\dag}(\vec
y).}
\medskip}

Indeed, using Eq.\bacaha\ and the canonical commutator Eq.\cancomm\
it follows immediately that the commutation relation satisfied by the $\phi_0$%
fields is
\eqn\gracomm
{\phi_0^{\dag}(\vec x)\phi_0^{\dag}(\vec y)-e^{\pm i \pi s}
\phi_0^{\dag}(\vec y)\phi_0^{\dag}(\vec x)=0,}
where the minus (plus) sign applies if the fields are commuted by
clockwise (counterclockwise) rotation. The other commutators are modified in an
analogous way. Thus, choosing the value of  the parameter $s$, the
fields $\phi_0$, $\pi_0$ can be made to satisfy anticommutators, or
generalized commutators (usually referred to as graded commutators).
It would thus appear that after eliminating the
Chern-Simons coupling, the field operators have acquired generic statistics

\bigskip
\noindent{\it Problems and paradoxes}
\smallskip
Despite its simplicity, the construction of field operators which satisfy
generalized commutation relations and carry generalized statistics which we
have just
presented runs into several problems and paradoxes. In particular, a closer
look to the derivation which has led to Eq.\gracomm\ reveals a physical
problem, and a mathematical paradox.
 Let us discuss them in turn.

Physically, if the operators $\phi_0$ are to be interpreted as field operators
with fractional statistics, we would expect them also to create excitations
with fractional spin. This must be true in the specific case of fermions:
if we choose the value of $s$ so that
Eq.\gracomm\ gives an anticommutator, then if the field
$\phi_0^\dagger$ is a {\it bona fide} fermi field operator, when acting
on the vacuum it must
create a state which carries half-integer spin.
This implies that if we rotate
such a state by $2\pi$, we must get the same state multiplied by the phase
$-1$..

Now, the states $\phi_0$ do have peculiar rotational properties because
the function $S(\vec x)$ Eq.\puregau\ is not rotationally invariant (if we take
$\Theta$ to be a multivalued function):
\eqn\rotse
{R^{2\pi}S(\vec x)=S(\vec x)+2\pi \int\! d^2x\, j^0(\vec x),}
which implies that
\eqn\rotph
{\eqalign{R^{2\pi}\phi_0^\dagger(\vec x)|
0\rangle&= e^{i 2\pi \int\! d^2x\, j^0(\vec x)}
\phi_0^\dagger(\vec x)|0\rangle\cr
&=e^{i 2\pi s}\phi_0^\dagger(\vec x)|0\rangle\cr},}
where in the last step we have used the commutator Eq.\crop, \ie, the fact that
$\phi^\dagger$ creates one unit of charge.

But comparing the graded commutator Eq.\gracomm\ with the transformation law of
the fields Eq.\rotph\ it  is immediately clear that the correct spin-statistics
relation is not compatible with the interpretation of the redefined field
operators $\phi_0$ as creation operators for field with generalized statistics,
even in the case of fermions. Indeed, if we take $s=1$, then Eq.\gracomm\ gives
an anticommutator, but the field $\phi_0$ is invariant upon rotation by $2\pi$.
If we take $s={1\over 2}$ then the field acquires a factor $-1$ upon
$2\pi$-rotation, as a fermion should, but Eq.\gracomm\ is no longer an
anticommutator. Thus there is no way we can obtain from $\psi_0$
simultaneously the good rotational and commutation relations of a fermion
field, let alone  those of a field with generic statistics. It seems that the
construction of $\phi_0$ as field operators with fractional spin and statistics
is physically inconsistent\ref\forjo{This problem with the approach of
Ref.\semen\ is pointed out by Forte and Jolic\oe ur, Ref.\csref.}.

There is also a mathematical problem in the previous derivation. The
derivation
was based on the fact that the expression
Eq.\gausol\ for $A^i$  satisfies Eq.\seccon. Let us check this explicitly.
The expression Eq.\twolap\ of the Green function of the two-dimensional
Laplacian, taken jointly with the property Eq.\exer\ of the function
$\Theta(\vec x)$ implies
\eqn\thedel
{\epsilon^{ij}\partial_i\partial_j \Theta(\vec x)=2\pi \delta^{(2)}(\vec x),}
which even though unusual is not contradictory
because $\Theta(\vec x)$ is clearly
ill-defined in the origin, thus derivatives acting on it may not commute in
that point. Then, using Eq.\thedel\ in the expression Eq.\puregau\ of $A^i$
we indeed get
\eqn\conchk
{\epsilon^{ij}\partial_iA_j(\vec x)=s\int\!d^2y\,\epsilon^{ij}
\partial_i\partial_j \Theta(\vec x-\vec y)j^0(\vec y,t)=2\pi s j^0(\vec x),}
consistently with the claim that $A^i$ Eq.\gausol\ solves the constraint
Eq.\seccon.

However, we may compute Eq.\conchk\ in an alternate, presumably equivalent way:
we change the integration variable $\vec y\to \vec x-\vec y$ in the definition
of $S(\vec x)$, Eq.\puregau. Then we get
\eqn\conchka
{\epsilon^{ij}\partial_iA_j(\vec x)=s\int\!d^2y\,\epsilon^{ij}
\partial_i\partial_j\left[ \Theta(\vec y)j^0(\vec x-\vec y,t)\right]=0,}
if the charge density is a smooth function. Thus we arrive at a contradiction,
unless the charge density vanishes or is singular. It seems thus that the
construction is also mathematically inconsistent.\ref\jacpi{This
apparent paradox in
the approach of Ref.\semen\  is pointed out (and resolved, see below)
by R.~Jackiw and S.~Y.~Pi,
{\it Phys. Rev.} {\bf D 42} 2500 (1990).}

These two problems point towards some deeper difficulties in extending the
construction of fractional spin and statistics from first-quantized
point-particle mechanics
to second-quantized field theory:\hfill\break
a) In quantum mechanics we have seen that it is possible to redefine the wave
functions in such a way that their transformation properties upon, say,
rotations are modified, while the spectrum of physical observables, such as
spin and statistics, is unchanged. In order to establish the
spectrum of physical observables knowledge of the interaction dynamics, and not
only the state vectors, is required. This suggests that knowledge of the
transformation properties of the field operators is not enough to determine
whether the spin and statistics of the fundamental excitations of the theory.
The first problem above suggests indeed that a naive identification of rescaled
field operators such
as $\psi_0^\dagger$ Eq.\gtran\ with creation operators is incorrect.
\hfil\break
b) In particle mechanics unusual transformation properties upon rotations and
boosts are induced on physical states through coupling to a superficially
rotationally invariant action (the Chern-Simons action) thanks to the
divergence of the particle-particle interaction at the singular points where
two particles coincide.
In field theory, the current $j^\mu(x)$
which carries the physical excitations is a smooth function over space-time,
hence a covariant interaction Lagrangian cannot consistently change the
transformation laws of smooth fields. \hfill\break
c) In field theory, the spin carried by a physical state ought to depend on its
particle content. In the approaches pursued so far, spin is induced on physical
state by means of a peculiar interaction. Whereas in the first-quantized theory
the form of the interaction depends on the number of particles, in the second
quantized theory the interaction is fixed and it is unclear how it can affect
physical quantum numbers in a way which depends on the
particle content.

Even though a systematic treatment is not yet available, we discuss a framework
where at least these problem find a satisfactory
answer\ref\stethi{The approach presented below
was proposed by Forte and Jolic\oe ur, Ref.\csref; see also Forte,
Ref.\jaccoc.}.
\bigskip

\subsec{\bf The operator cocycle approach}

We wish to pursue the same logic which has lead to fractional spin and
statistics in the first-quantized theory.
 To this purpose,
we need to generalize suitably the concepts of wave function and propagator.
This may be done in the Schr\"odinger functional formulation of
field theory.
Time is singled out, and the fields $\phi(\vec x)$ are quantized
canonically at fixed $t$.
The state vectors are then  functionals
of the field configurations, and functions of time, while the fields play the
role of coordinates:
\eqn\wfunct
{\langle q,t|\Psi\rangle=
\langle\phi(\vec x),t|\Psi\rangle=\Psi [\phi (x);t].}
The state functionals Eq.\wfunct\ are propagated by
\eqn\funprop
{K(\phi^\prime (\vec x),t^\prime ;\phi (\vec x),t)=
\int \!{\cal D}\phi (\vec x,t_0)\, \exp \left( i\int^{t^\prime}_t\!
dt_0\,\int d\vec x{\cal L}[\phi (\vec x,t_0)]\right),}
where $\cal L$ is the Lagrangian of the given field theory,
and the boundary conditions are the field configurations
$\phi (\vec x)$ at initial time $t$ and
$\phi^\prime (\vec x)$ at final time $t^\prime$.
Even though this procedure is not manifestly relativistically invariant
at intermediate stages, physical amplitudes ($S$-matrix
elements) are. Rather, the state functionals transform covariantly: upon
a Lorentz boost that takes the vector $\hat t$ into the time-like vector
$\hat n$ the physical states are transformed into functionals of
the fields quantized on the plane orthogonal to $\hat n$, at fixed values
of the coordinate along $\hat n$. In general, one may choose to quantize
the system canonically on a space-like plane $\Sigma$ and take the coordinate
orthogonal to $\Sigma$ to parametrize its evolution.

Then, suppose we start with a theory of bosons with Lagrangian
${\cal L}_0$, and we add to it
a topological Lagrangian  ${\cal L}_t$, \ie,
a Lagrangian which may be expressed as the
total divergence of a three-vector density:
\eqn\toplag
{\eqalign{{\cal L}_0&={\cal L}+{\cal L}_t\cr
{\cal L}_t&=\partial_\mu\Omega^\mu(x).}}
If we demand that fields fall
off at infinity, this leads to nonvanishing contributions at initial and
final times only, since there the field configuration is nontrivial because
of the boundary conditions:
\eqn\surter
{\eqalign{&\int\,d\vec xdt_0\,\partial_\mu\Omega^\mu[\phi (\vec x, t_0)]=
H(t^\prime)-
H(t)\cr
&\quad H(t)=\int \! d\vec x\, \Omega_0(\vec x, t).\cr}}

But then we can proceed exactly as in particle mechanics: we note  that the
propagator of the theory with Lagrangian $\cal L$ Eq.\toplag\ is related by
\eqn\fullprop
{K(q^\prime ,t^\prime ; q,t)=
e^{iH(t^\prime)}
K_0(q^\prime ,t^\prime ; q,t)e^{-iH(t)}}
to the propagator $K_0(q^\prime ,t^\prime ; q,t)$ of the theory with Lagrangian
${\cal L}_0$.
Again, the state functionals may be redefined according to
\eqn\redef
{\Psi_0[\phi(\vec x),t]=e^{-iH(t)}\Psi[\phi(\vec x),t],}
and then it follows that the $S$ matrix elements computed acting on
the
states $\Psi$ with the propagator Eq.\fullprop\ is the same as that found
acting with the propagator $K_0$ free of topological interaction on the
redefined states \redef.

However, in a quantized field theory the fields are operators. Hence, the
phase prefactor which enters the redefinition Eq.\redef\ of the state
vectors is itself an operator, whose effect on the physical states will in
general depend on their particle content. Furthermore, the quantized field
operators will display short-distance divergencies which may take the place of
the divergencies which appeared in the first-quantized theory when two
particles came to coincide. We shall take advantage of these facts to solve the
problems discussed above.

\bigskip
\noindent{\it The Hopf interaction in field theory}
\smallskip
An obvious guess for a candidate
topological Lagrangian Eq.\toplag\ is the Hopf current-current
interaction Eq.\hac, which we can write in any
theory which admits a conserved current $j^\mu$. Using the same trick as in
Sect.II.2, namely rewriting the bilocal kernel in the Hopf interaction
Eq.\biker\ in terms of a monopole potential, Eq.\dirmon, we may cast the Hopf
action in the form
\eqn\hoptot
{\eqalign{I_H &=-{s\over2}\int\! d^3 x\, d^3 y\, j^\mu (x)\left[
\partial_\mu \tilde A_\nu (x-y) -
\partial_\nu \tilde A_\mu (x-y) \right] j^\nu (y)\cr
&=-{s\over 2}\int\!d^3x\,\partial_\mu\Omega^\mu,\cr}
}
where
\eqn\exsurf
{
\Omega_\mu(x)=j^\mu(x) \int\! d^3 y\, \left(\tilde A_\rho (x-y) j^\rho (y)+
\tilde A_\rho (y-x) j^\rho (y)\right).}

This is not quite topological,
because $\Omega_\mu(x)$ Eq.\exsurf\ is still nonlocal in time
(it  is defined
as an integral over all times).
However, the divergent nature of the
monopole potential when $x\to y$ allows a local determination
of the surface terms according to Eq.\surter.
An almost verbatim rerun of the computation which lead from the Hopf
interaction Eq.\finiij\ to the winding number Eq.\resiij\ leads, in field
theory, to
\eqn\ftresh
{I_t=  -2s\left[H(t^\prime)-H(t)\right]+I_{\rm cov} }
where the surface term is explicitly given by
\eqn\ftsurh
{\quad H(t)={1\over2}
\int\!d^2x\,d^2y\,j^0(\vec x;t)\Theta(\vec x-\vec y)j^0(\vec y;t),}
and $I_{\rm cov}$ denotes a contribution which cannot be simply cast as a
surface term, but is covariant upon Lorentz transformation, \ie, it is the
field theoretic analogue of the terms $I_g$ found in particle mechanics
[Eq.\resiij]. The
result Eq.\ftresh\ is derived under the only assumption that the current
$j^\mu$ be conserved as a symmetry (Noether) current (so that its conservation
holds at the quantum level, as expressed by Ward identities).

 The function $H(t)$ Eq.\ftsurh\
may seem at first quite ill-defined, since the function $\Theta(\vec x)$ is
ill-defined when $|\vec x|\to 0$. Indeed, at the classical level $I(t)$
Eq.\ftresh\
is Lorentz invariant, implying that $H(t)$
is a rotationally invariant and Lorentz covariant quantity. However,
if we use a point-particle expression (\ie\ the usual sum of deltas
Eq.\curr) for the charge densities in Eq.\ftresh\
then $H(t)$ reduces to the point-particle result Eq.\resiij, which is
manifestly rotationally noninvariant. This is at the root of  the
mathematical difficulties discussed at the end of Sect.V.2.
However, in quantum field theory the
propagation kernel Eq.\fullprop\ is an operator,
a functional of the field operators on
which the currents $j^\mu$ depend. The phases $e^{2siH(t)}$
Eq.\redef\ induced
on the state functionals are thus indeed to be viewed as operator-valued
quantities.
The fact that the bilocal kernel in $H(t)$  is ill-defined
at $x=y$ is then irrelevant because the product of the two charge densities
diverges when their arguments coincide as $j^0(x)j^0(y)\mathop{
\sim}\limits_{x\to
y}{1\over |x-y|^4}$. This point is thereby effectively excluded from the
integration domain in Eq.\ftsurh.

Rather than going through the technically involved checks that this is
enough to make $H(t)$ Eq.\ftresh\ well-defined,
let us discuss what is the effect of using the form Eq.\ftsurh\
of the phase induced by the topological interaction in the redefinition of
physical states according to Eq.\redef. Then we shall see how the two
problems discussed at the end of Sect.V.1 are  resolved.

\bigskip
\noindent{\it Operator cocycle and physical states}
\smallskip
An $n$-particle state
\eqn\npst
{\Psi^n[\phi(\vec x),t]=
\phi^\dagger(\vec x_1)\dots\phi^\dagger(\vec x_n)|0\rangle}
will lead to the redefined state Eq.\redef\
\eqn\rnpst
{\Psi_0[\phi(\vec x),t]=e^{2isH(t)}\phi^\dagger(\vec x_1)\dots
\phi^\dagger(\vec x_n)|0\rangle.}
Now, whenever the fields which are integrated over in the expression of $H(t)$
come close to the points $x_1,\dots,x_n$ there will be short-distance
divergences, which lead to the sought-for singularities. Thus, the phase
prefactor $e^{2isH(t)}$ should be viewed as an operator-valued cocycle, namely,
as an operator which, acting on physical states, produces as its eigenvalues
the cocycles appropriate to the various states. These can be computed by
performing an operator-product expansion in order to extract the leading
singularities in Eq.\rnpst.

Again, rather than following this procedure, we deduce
the same result
by
somewhat formal, even though much simpler manipulations. We make use
of the commutation relation
\eqn\hcomm
{[H(t),\phi^{\dag}(\vec z,t)]=
S(\vec z,t)\phi^{\dag}(\vec z,t).}
This  follows from the assumption that there exists a creation operator
$\phi^\dagger$ which satisfies the basic commutator Eq.\crop.
\bigskip
{\narrower\narrower \noindent{\bf Exercise}: Prove that
\eqn\npartph
{\eqalign{e^{2isH}|\Psi_0^n\rangle&=
\prod_{i=1}^n \left(e^{2isS(\vec x_i)}\phi^{\dag}(\vec x_i)\right)
\widetilde{|0\rangle}
\cr
&=
e^{-2is
\sum_{j=1}^n\sum_{i=1}^{j-1}
\Theta(\vec x_i-\vec x_j)}\left[e^{2is\sum_{i=1}^n S(\vec x_i)}
\prod_{i=1}^n \phi^{\dag}(\vec x_i)|\widetilde{0\rangle}\right], \cr }}
where $\Psi_0^n$ is given by Eq.\rnpst\ and
\eqn\revac
{\widetilde{|0\rangle}=e^{2isH}|0\rangle.}
\medskip}

\noindent
Then, the redefined one- and two-particle state functionals are
\eqn\onantw
{\eqalign{
\Psi^1_0=e^{2isH}\phi^{\dag}(\vec x;t)|0\rangle&=
e^{2isS(\vec x)}\phi^{\dag}(\vec x;t)\widetilde{|0\rangle}
\cr
\Psi^2_0=e^{2isH}\phi^{\dag}(\vec x;t)\phi^{\dag}(\vec y;t)|0\rangle&=
e^{2is\left[S(\vec x)+S(\vec y)\right]}e^{-2is\Theta(\vec x-\vec y)}
\phi^{\dag}(\vec x;t)\phi^{\dag}(\vec y;t)\widetilde{|0\rangle},\cr}}
and so forth. Here $\widetilde{|0\rangle}$
is a redefined vacuum,
which is generally
different from $|0\rangle$ because even though $Q|0\rangle=0$, in general
$j^0(\vec x)|0\rangle\not=0$. However, the redefinition does not affect the
Poincar\'e invariance of the vacuum, and amounts to normal ordering.
Hence, the operator cocycle provides phase prefactors both on one- and
many-particle states, while leaving the vacuum invariant (up to normal
ordering).

The many-particle states, however, may still be symmetrized: namely,
we are free to choose the
symmetry of physical states by symmetrizing the states on which the operator
phase $e^{2isH}$ acts, so that the general two-particle (say)
state will be
\eqn\twops
{\Psi^2_0=
e^{2is\left[S(\vec x)+S(\vec y)\right]}e^{i\sigma\Theta(\vec x-\vec y)}
\phi^{\dag}(\vec x;t)\phi^{\dag}(\vec y;t)\widetilde{|0\rangle},}
where $\sigma$ is a free parameter.
Now, it is clear that the parameters $s$ and $\sigma$ control
respectively the spin and
statistics of $n$-particle states. Indeed, the latter coincides with the
statistics as defined in Eq.\st, because the $\Theta$ dependent phases
in Eq.\twops\ just symmetrize the state with respect to the interchange of the
quantum numbers $x_1$, $x_2$. This also leads to a
contribution to angular momentum, due to the (kinematical) relation between
statistics and angular momentum discussed in the end of Sect.IV.2, according to
Eq.s~\ssta.
The former is related to spin because of the transformation law of $S(\vec x)$,
Eq.\rotse.
Also,  that Eq. shows that the variation of
the function $S(x)$ upon rotations, \ie\ its contribution  to spin will depend
on the charge of the state. In
particular for an $n$ particle state, if there are $n$
phase prefactors:
\eqn\rotph
{R^{2\pi}e^{2is\left[S(\vec x_1)+\dots+S(\vec x_n)\right]}
\phi^\dagger(\vec x_1)\dots\phi^\dagger(\vec x_n)|0\rangle
=e^{2isn^2}
\phi^\dagger(\vec x_1)\dots\phi^\dagger(\vec x_n)|0\rangle,}
\ie, the dependence of the spin on the number of particles is quadratic.

It may be furthermore shown, through straightforward generalization of the
techniques discussed in Sect.III.1, that upon generic Lorentz transformations
the phase prefactors $S(x)$ and $\Theta(x)$ in Eq.\twops\ transform with the
correct Lorentz and Poincar\'e cocycles appropriate to their particle content.
The spectrum of spin, orbital and total angular
momentum is thus found to be equal to
\eqn\ftspectra
{\eqalign{
\ell&=n(n-1)\sigma+k;\quad k\in\Z\cr
S&=n^2s^\prime\cr
j&=\ell+S,\cr}}
where $s^\prime=2s$,
to be contrasted with the point-particle results, Eq.\spectra.

Comparison of the point particle and field theoretical angular momentum
spectra,
Eq.s~\spectra\ and \ftspectra\
shows that: i) the dependence of the spin and statistics on
the coefficient of the topological action is by a factor of 2 larger in the
field theory; ii) the statistics is a free parameter in the field theory while
it is fixed a priori in the particle theory; iii) the dependence of the
statistics on the number of particles is the same while that of the spin isn't.
This means that the second quantization of the theory does not
commute with the point particle limit, and can be traced to the different way
the repulsive core which gives rise to fractional spin and statistics is
treated. Namely, in both case the repulsive interaction which gives rise to a
multiply connected configuration space (required for fractional spin) and the
exclusion principle (required for fractional statistics) is due to the
divergence of the bilocal kernel Eq.\biker\  as its arguments $x$, $y$
coincide.
This divergence however is regulated differently. In particle mechanics it is
regulated geometrically, by evaluating the kernel over particle trajectories,
which leads to the regular integrand of Eq.\hopsum. In field theory it is
regulated by the current-current repulsion due to their short-distance
divergencies.

We notice finally that all the results derived here have been obtained without
explicitly specifying the Lagrangian of the theory
whose conserved current is coupled through the Hopf interaction, and
rely only on the
assumption that a conserved current and  a creation operator satisfying
Eq.\crop\ exist. Even though this includes a large class of theories, it is
worth pointing out that the Lagrangian Eq.\ftcs\ discussed in Sect.V.1 is not
in this class: indeed, in that theory the conserved current is not a symmetry
current of the theory defined without Chern-Simons interaction, but rather a
dynamically conserved current which depends on the Chern-Simons field.

\bigskip
\noindent{\it Resolution of the paradoxes}
\smallskip
Even without getting into the details of the operator cocycle approach we can
see how the paradoxes discussed at the end of Sect.V.2 are resolved\ref\resol{
The arguments given in
the sequel are based on Forte and Jolic\oe ur, Ref.\csref.
A totally different resolution, within a different approach to field theory
with fractional statistics,
has been presented by A.~Liguori, M.~Mintchev and M.~Rossi, {\it Phys. Lett.}
{\bf B303}, 38; {\bf B305}, 52 (1993). The relation of this approach to that
discussed here has not been investigated.}.
The spin-statistics paradox requires us to take a closer look at the
spin-statistics relation in the present approach.
Eq.\ftspectra\
shows that if we require the field theory to be local and well-defined
in the thermodynamic limit then a particular spin-statistics relation
is singled out.
Indeed, noninteracting in and out states can exist only if the total angular
momentum (which is an additive quantum number, because the rotation group is
abelian) is linear in the number of particles. Otherwise, either noninteracting
states do not exist, in which case the thermodynamic limit is ill-defined, or
causality is violated.
To see that  this is true,
perform the following {\it Gedankenexperiment}. First, measure
the spectrum of $j$ for a localized
system of $n$ particles; then, add a particle to the
system and measure the spectrum again. If $j$ isn't linear in $n$ the
difference in normalization of the two spectra depends on the total number
of particles which are arbitrarily far and causally disconnected from the
system, \ie, it depends on the ``wave function of the universe''.

The requirement of linearity of $j$ Eq.\ftspectra\ in the number
of particles $n$ is satisfied if
\eqn\ftssta
{s^\prime=-\sigma}
which implies
\eqn\totspec
{j=k+ns^\prime \quad k\in\Z.}
This is a genuine spin-statistics theorem; it has the opposite
sign as that which one might have been naively guessed, and which is displayed
by the point particle theory, Eq.\spectra.
However, if spin is
integer or half-integer
Eq.\ftspectra\ reduces to the usual relation and there is no
difference between the field theory and the point particle case.

In order that the spin-statistics  relation Eq.\ftssta\
be satisfied, a nontrivial symmetry has to be imposed on
physical states, \ie, the statistics must differ from that automatically
generated by the operator phase, and displayed in Eq.\npartph.
This prevents the
identification of redefined  operators as creation
operators for particles with fractional spin, because an extra symmetrization
is required after the creation operators and the operator cocycle have been
applied. Thus this analysis shows that
creation and annihilation operators for generic
statistics cannot be identified, yet the
spin-statistics theorem is satisfied (and has a nontrivial generalization to
the
fractional case). This resolves the spin-statistics paradox.

The paradox in the manipulations using the purported expression of the
Chern-Simons field as a pure gauge, Eq.\puregau\ requires us to specify more
carefully the integration domains in the definitions of the functions $H(t)$
Eq.\ftsurh\ and $S(\vec x)$ Eq.\puregau. Indeed, a careful analysis of the
computation which leads from Eq.\exsurf\ to the expression Eq.\ftsurh\ of the
surface terms generated by the Hopf action reveals that the extremes of
integration in the definition of $H(t)$ read
\eqn\careh
{\eqalign{H(t)=&{1\over2}\int_0^\infty\!\rho_x
d\rho_x\,\int_\alpha^{\alpha+2\pi}\!
d\theta_x\,
\int_{0}^{\infty}\!\rho
d\rho\,\int_{\theta_x}^{\theta_x+2\pi}\!d\theta\,\cr
&\qquad\quad \theta\left[j^0(\vec x ,t )
j^0 (\vec x+\vec r,t)\right] ,\cr}}
where $(\rho_x,\theta_x)$ and $(\rho,\theta)$ are polar components of
the vectors $\vec x$, $\vec r$, respectively,
and $\alpha$ is an arbitrary (multivalued) reference angle,
which may be chosen, as usual, by defining $H(t)$ as
the integral of its time derivative from a reference field
configuration to the given one.

This means that the precise definition of $S(\vec x)$ is
\eqn\cares
{S(\vec x)=\int_{\theta_x}^{\theta_x+2\pi}\!d\theta\,\int_0^\infty\!dr r\,
\theta j^0(\vec x+\vec y^\prime,t),}
where $\vec y^\prime=(r\cos\theta,r\sin\theta)$.
Hence, the function $\Theta$ in the definition of $S(\vec x)$ is a {\it
multivalued} function of the polar component $\theta_x$ of the vector $\vec x$
on which $S$ depends, but it is a single-valued function in the integration
domain with respect to $\vec y$.
\bigskip
{\narrower\narrower \noindent{\bf Exercise}: Prove that $S(\vec x)$ defined by
Eq.\cares\ is multivalued upon rotations according to Eq.\rotse, by computing
the action of the angular momentum operator
$L=-i\epsilon^{ab}x^a\partial_b$ on it and showing that
\eqn\rotcare
{\eqalign{L iS(\vec y)&=\int d^2y \epsilon^{ab}x^a\left(-\epsilon^{bc}\right)
{\left(x-y\right)^c\over\left|\vec x-\vec y\right|^2} j^0(\vec y)
+\int_{|y|>|x|}\!\!\!\!\!\! d^2 y\, j^0(y)\cr
&\quad=\int d^2 y j^0(\vec y).\cr}}
Assume for simplicity that
$j^0=$ is
rotationally invariant.
\medskip}

The paradox described above then disappears because the crucial relation
Eq.\thedel\ no longer holds. Indeed, Eq.\cares\ shows that
in the definition Eq.\puregau\ the integration
over $\vec y$ is to be performed by taking a definition of $\Theta$ which
has a branch cut along $\theta_x$, \ie, $\theta_x\le\Theta(\vec x-\vec y)
\le\theta_x+2\pi$. But if $\Theta$ has a branch cut, then along the cut
the basic property of $\Theta$, Eq.\exer, is modified because of the
discontinuity in the value of $\Theta$ along the cut.
If, \eg, the cut is along the positive
$x$-axis then Eq.\exer\ is modified to
\eqn\cutth
{{\partial \over \partial x^a}\Theta(\vec x) =
-\epsilon_{ab}{x^a \over |\vec x|^2}
\,-2\pi H(x^1 )\delta (x^2)\left[\matrix{0\cr1\cr}\right]_a,}
where $H$ is the Heaviside step function.\foot{which we refrain from calling as
usual $\Theta$ for obvious reasons}
But from Eq.\cutth\ it follows that
\eqn\ntd
{\epsilon^{ij}\partial_i\partial_j \Theta(\vec x)=0,}
thus the paradox disappears.

Equivalently, the definition Eq.\cares\ of $S(\vec x)$ may be viewed
as the result of taking a multivalued determination of $\Theta$, and
extending the integration over the full Riemann surface, but with a
particle density which is nonvanishing only between
$\theta_x$ and $\theta_x+2\pi$.
\bigskip
{\narrower\narrower \noindent{\bf Exercise}: Assume
that in the definition of
$S(\vec x)$ Eq.\puregau\ the integration is extended on the full Riemann
surface of the logarithm, \ie, $\-\infty\le\Theta(\vec x - \vec y)\le\infty$,
but $j^0$ is nonvanishing only in a range of values
$\theta_0\le \Theta(\vec x - \vec y)<\theta_0+2\pi$. Show that under such
assumptions
the interchange of
integral and derivative in Eq.\puregau\ fails. Compute the correction and show
that (assuming for simplicity $\theta_0=0$)
it is given by\ref\jacpriv{This argument was developed by R.~Jackiw
({\it private communication}), see also Jackiw and Pi, Ref.\jacpi.}
\eqn\jachag
{ A^i(\vec x,t)-s\partial_iS(\vec x)-2\pi\int_{-\infty}^x\!dx^\prime\,
j^0(x^\prime,y,t).}
\medskip}

\noindent
In this case
Eq.\exer\ holds, but because of Eq.\jachag\ $S(\vec x)$ cannot be written as a
pure gauge; derivative and integral do not commute and the non-commutativity
resolves the discrepancy between Eq.\conchk\ and Eq.\conchka.
\topinsert
\vskip 7.5truecm
\figcaptr{3:}{a) A few sheets of the Riemann surface of the logarithm where the
function $\Theta(\vec x-\vec y)$ of Eq.\cares\ is defined. b) The charge
density
as viewed from the creation operator $\phi^\dagger(\vec x)$ is nonzero only
in the
shaded area. c) After time evolution the area occupied by the charge
density has moved.
\smallskip}
\endinsert

The definition of $S$ Eq.\cares\ may seem awkward, but it is actually
the physically natural generalization of the definition adopted in point
particle theory. Recall that in that case $\Theta$ was viewed as a multivalued
relative polar angle between all couples of particles. This function takes
values on the cover of the two-particle relative space, which is the
infinitely-sheeted Riemann
surface of the logarithm.
In Eq.\cares\ $S$ is
also the multivalued relative polar angle,
weighted by the particle density. Because,
however, the particle density is defined on configuration space, the
integration is defined on the configuration space, rather than on the
infinitely sheeted surface, whereas the multivaluedness is contained in the
determination of the relative angle. Equivalently, the charge density as viewed
by the creation operator\foot{Notice that
the dependence of $S(\vec x)$
on $\vec x$ is due to the commutation relation Eq.\hcomm\ with
$\phi^\dagger(\vec x)$, \ie\ it is induced by the creation operator.}
$\phi^\dagger(x)$
occupies a $2\pi$ range on the infinitely-sheeted
Riemann surface, along which it may move in the course of the time evolution of
the system (see Fig.3).

\vfill
\eject
\newsec{\bf OPEN PROBLEMS}

In these lectures we have discussed an approach to the quantization of systems
with fractional spin and statistics which is amenable to a relativistic
treatment, and seems to carry through from quantum mechanics to field theory.
The approach is based on the path-integral quantization of spin using
phase-space variables. Even though this approach has allowed us to give a
complete and consistent description of the kinematics of these systems,
dynamical results are conspicuously missing. However elegant the mathematics
involved, we have only proven that the particle content of the theories we have
studied is consistent with the desired transformation properties under the
Lorentz and Poincar\'e groups.
Nevertheless, our analysis has led to at least one nontrivial result, namely
that the spin-statistics relation seems to have a different structure in
quantum mechanics and field theory.

The first steps we described in the formulation of relativistic quantum
mechanics should hopefully open the way to an investigation of dynamical
problems, such as bound states problems, scattering, electromagnetic coupling.
This should also shed light on the relationship
between the cocycle approach, described here, and
the alternative approach based on infinite-component wave functions.
The construction of a satisfactory field theory with fractional statistics
still has a long way to go: we would like, first of all, to understand the
structure of the Hilbert space for such a theory. This should allow one to
reach a deeper understanding of spin-statistics relations, and eventually of
the relativistic statistical mechanics of such systems.
Perhaps, the intricacies we found are hinting to new physical structures which
are awaiting discovery.

\bigskip
\centerline{\bf ACKNOWLEDGEMENTS}
I am grateful to Oscar Eboli and Victor
Rivelles for inviting me to give these lectures, and
to all the organizers for providing a relaxed yet stimulating
atmosphere during the school. I thank all those who attended the lectures for
their interest and especially K.~Lechner for several discussions.
\vfill
\eject
\listrefs
\vfill\eject
\bye